\documentclass[useAMS,usenatbib]{mn2e}

\usepackage{aas_macros}

\usepackage{graphicx}
\title[Spiral galaxy structure in STAGES]{The environmental dependence of the structure of outer galactic
discs in STAGES spiral galaxies}

\author[D.~T.~Maltby et al.]
{David~T.~Maltby$^{1}$\thanks{E-mail: ppxdtm@nottingham.ac.uk},
 Meghan~E.~Gray$^{1}$,
 Alfonso~Arag{\'o}n-Salamanca$^{1}$,
 Christian~Wolf$^{2}$,
 \newauthor Eric~F.~Bell$^{3}$,
 Shardha~Jogee$^{4}$,
 Boris~H{\"{a}}u{\ss}ler$^{1}$,
 Fabio~D.~Barazza$^{5}$,
 Asmus~B\"{o}hm$^{6}$,
\newauthor Knud~Jahnke$^{7}$.\\
$^{1}$School of Physics and Astronomy, The University of Nottingham, University Park, Nottingham, NG7 2RD, UK. \\
$^{2}$Department of Physics, Denys Wilkinson Building, University of Oxford, Keble Road, Oxford, OX1 3RH, UK. \\
$^{3}$Department of Astronomy, University of Michigan, 830 Dennison Building, 500 Church St., Ann Arbor, MI 48109, USA. \\
$^{4}$Department of Astronomy, University of Texas at Austin, 1 University Station, C1400 Austin, TX 78712-0259, USA. \\
$^{5}$Department of Physics, University of Basel, Klingelbergstrasse 82, 4056 Basel, Switzerland. \\
$^{6}$Institute of Astro- and Particle Physics, University of Innsbruck, Technikerstr. 25/8, A-6020 Innsbruck, Austria. \\
$^{7}$Max-Plank-Institut f{\"{u}}r Astronomie, K{\"{o}}nigstuhl 17, D-69117, Heidelberg, Germany.} 

\begin{document}

\date{Accepted Year Month Date. Received Year Month Date; in original form Year Month Date}

\pagerange{\pageref{firstpage}--\pageref{lastpage}} \pubyear{0000}

\maketitle

\label{firstpage}


\begin{abstract}

We present an analysis of $V$-band radial surface brightness profiles for spiral galaxies from the field and
cluster environments using {\em Hubble Space Telescope}/Advanced Camera for Surveys imaging and data from
the Space Telescope A901/2 Galaxy Evolution Survey (STAGES). We use a large sample of $\sim 330$ face-on to
intermediately inclined spiral galaxies and assess the effect of the galaxy environment on the azimuthally
averaged radial surface brightness $\mu$ profiles for each galaxy in the outer stellar disc ($24 < \mu <
26.5\,{\rm mag\,arcsec}^{-2}$). For galaxies with a purely exponential outer disc ($\sim50$ per
cent), we determine the significance of an environmental dependence on the outer disc scalelength $h_{\rm
out}$. For galaxies with a broken exponential in their outer disc, either down-bending ({\em
truncation}, $\sim10$ per cent) or up-bending ({\em anti-truncation}, $\sim40$ per cent), we measure the
strength $T$ (outer-to-inner scalelength ratio, ${\rm log}_{10}\,h_{\rm out}/h_{\rm in}$) of the $\mu$
breaks and determine the significance of an environmental dependence on break strength $T$. Surprisingly, we
find no evidence to suggest any such environmental dependence on either outer disc scalelength $h_{\rm out}$
or break strength $T$, implying that the galaxy environment is not affecting the stellar distribution in the
outer stellar disc. We also find that for galaxies with small effective radii ($r_e < 3\,\rm kpc$) there is
a lack of outer disc truncations in both the field and cluster environments. Our results suggest that the
stellar distribution in the outer disc of spiral galaxies is not significantly affected by the galaxy
environment.

\end{abstract}

\begin{keywords}

galaxies: clusters: general ---
galaxies: evolution ---
galaxies: spiral ---
galaxies: structure ---

\end{keywords}

\section[]{Introduction}

\label{Introduction}

The structure of the outer regions of galactic discs is important to our understanding of the formation and
evolution of spiral galaxies. These faint, outer regions are more easily affected by interactions with other
galaxies, and therefore their structural characteristics must be closely related to their evolutionary
history. Certain physical processes inherent to galaxy evolution and dependent on the galaxy environment,
for example, ram pressure stripping of the interstellar medium, mergers, and harassment \citep[e.g.,
][]{Gunn_Gott:1972, Okamoto_Nagashima:2004, Moore_etal:1996} could have an effect on the galactic disc and
therefore the light distribution (surface brightness $\mu$ profile) of a galaxy. Therefore, the
environmental dependence of the shape of spiral galaxy radial surface brightness profiles in the outer
galactic disc will provide evidence for the physical processes of galaxy evolution occurring in different
environments.

Since the work of \cite{Patterson:1940}, \cite{deVaucouleurs:1959}, and \cite{Freeman:1970} we have known
that the light profiles of spiral galaxies are comprised of two main components: an inner component
dominated by a bulge, and an outer disc that follows a simple exponential decline with minor deviations
related to the spiral arms. However, this classical picture does not hold for most spiral galaxies in the
Universe and has been shown to fail at the faint surface brightness of the outer stellar disc
\citep{vanderKruit:1979, Pohlen_etal:2002}. Since \cite{vanderKruit:1979} we have known that the exponential
decline in the outer disc does not extend out to the last measured point but can be truncated (sharply cut off)
after several scalelengths. More recently \cite{Pohlen_etal:2002} have shown that the exponential decline
does not cut off completely at the truncation. They find that most profiles are actually best described by a
two slope model (broken exponential), characterised by an inner and outer exponential scalelength separated
by a relatively well defined break radius $r_{\rm brk}$. Many studies have now reported (mainly using surface
photometry) the existence of broken exponential stellar discs (truncations) in spiral galaxies in both the
local \citep{Pohlen_etal:2002, Pohlen_Trujillo:2006, Pohlen_etal:2007, vanderKruit:2007, Bakos_etal:2008}
and distant $z < 1$ Universe \citep{Perez:2004, Trujillo_Pohlen:2005, Azzollini_etal:2008}.

As a result of these studies it has been shown that disc galaxies can be classified into three broad types
(Type I, II, and III) according to break features in their radial stellar surface brightness profiles
\citep[see e.g.][]{Pohlen_Trujillo:2006}. Type I (no break), the galaxy has a simple exponential profile
extending out to several scalelengths \citep[e.g.][]{BlandHawthorn_etal:2005}. Type II (down-bending break,
{\em truncation}), the exponential is broken with a shallow inner and steeper outer exponential region
separated by a relatively well defined break radius \citep{vanderKruit:1979, Pohlen_etal:2002}. Type III
(up-bending break, {\em antitruncation}), a class recently discovered by \cite{Erwin_etal:2005} which have a
broken exponential with the opposite behaviour to a Type II profile. In each case the classification refers
to the outer, disc component of the galaxy radial surface brightness profile and does not consider the inner
varying bulge component. In some cases the inner bulge component may be near exponential in nature, however,
this classification scheme only considers the disc component and is fairly insensitive to the nature of the
inner (bulge) profile.

Measurements independent of surface photometry (from resolved star counts) are also available on nearby
galaxies for each of the three profile types. \cite{BlandHawthorn_etal:2005} find that NGC300 has a
simple exponential profile extending out to $\sim10$ scalelengths (Type I), \cite{Ferguson_etal:2007} argue
that M33 is best described as a broken exponential with a down-bending break (Type II), and
\cite{Ibata_etal:2005} report that M31 could be described as having an antitruncated disc (Type III).

The physical processes that cause the different types of profile are not well understood. Some models
suggest that Type II profiles (truncations) could be due to the effect of a star formation threshold in disc
column density towards the edge of the disc \citep[e.g.][]{Kennicutt:1989, Elmegreen_Parravano:1994,
Schaye:2004}. This theory may account for a sharp truncation (cut off), but does not explain observations of
extended outer exponential components in some galaxies \cite[e.g.][]{Pohlen_etal:2002}. Another theory has
been proposed by \cite{Debattista_etal:2006} who find down-bending breaks in simulated disc profiles solely
from collisionless N-body simulations. They suggest Type II galaxies are the consequence of a resonance
phenomenon and a redistribution of angular momentum that leads to an increased central density and surface
brightness. Many models now incorporate both these ideas and suggest that the inner disc forms as a
consequence of the star formation threshold while the outer disc forms by the outward migration of stars
from the inner disc to regions beyond the star formation threshold. This migration could be due to resonant
scattering with spiral arms (\citealt{Roskar_etal:2008a}a; \citealt{Roskar_etal:2008b}b) or clump disruptions
\citep{Bournaud_etal:2007}. Alternatively, another model is suggested by \cite{Foyle_etal:2008} who use
simulations to show that by starting from a single exponential disc, the inner disc forms as the bulge draws
mass from the inner regions. As a consequence the inner disc profile becomes shallower while the outer
region stays almost unaffected. Less work has been carried out on the origin of Type III galaxies, but
\cite{Younger_etal:2007} have shown that recent minor mergers could produce up-bending stellar profiles in
the remnant galaxy.

However, all these models rely on the break in the surface brightness profile having an analogous break in
the stellar surface mass density profile. \cite{Bakos_etal:2008} used colour profiles to calculate stellar
mass surface density profiles for a sample of Type II and Type III galaxies identified by
\cite{Pohlen_Trujillo:2006}. Interestingly, for Type II galaxies they find the stellar mass surface density
profiles are almost purely exponential. Therefore, the break in the surface brightness profile is not
necessarily related to a break in the stellar mass surface density profile. Consequently, the surface
brightness break may not be due to the distribution of stellar mass but could be due to a radial change in
the stellar population. For Type III galaxies however, they do find an analogous break in the stellar
surface mass density profile indicating these breaks could be due to the stellar mass distribution.

Investigating the frequency of profile types in different galaxy environments will provide evidence for
their origin and the effect of the environment on the outer stellar disc. Presently, there have been
few systematic searches for stellar disc truncations in spiral galaxies in either the local
\citep{Pohlen_Trujillo:2006} or distant Universe \citep{Trujillo_Pohlen:2005, Azzollini_etal:2008}. 
\cite{Pohlen_Trujillo:2006} use a sample of $\sim90$ face-on to intermediate inclined nearby late-type
(Sb-Sdm) spiral galaxies from the Sloan Digital Sky Survey \citep[SDSS;][]{York_etal:2000} and find that
approximately $10$ per cent are Type I, $60$ per cent are Type II and $30$ per cent are Type III. They also
report that the shape of the profiles correlate with Hubble type. In their sample down-bending breaks are
more frequent in later Hubble types while the fraction of up-bending breaks rises towards earlier types.
They also find no relation between the galaxy environment, as determined by the number of nearest neighbours,
and the shape of the surface brightness profile. However, they do not reach very dense environments and
low number statistics did not allow for major conclusions. 
\cite{Azzollini_etal:2008} recently conducted the largest systematic search for stellar disc truncations yet
undertaken at intermediate redshift ($0.1 < z < 1.1$) using the Great Observatories Origins Deep Survey
\citep[GOODS;][]{Giavalisco_etal:2004} south field. They use a sample of $505$ galaxies and obtain the
frequency of profile types in different redshift ranges. They find that the frequency of profile types
(Type I:II:III) is $25$:$59$:$15$ per cent for $0.1 < z < 0.5$, and does not vary significantly with
redshift out to $z\sim1.1$.

The aim of this paper is to undertake a systematic search for broken exponentials in the field and cluster
environment using the Space Telescope A901/2 Galaxy Evolution Survey \cite[STAGES;][]{Gray_etal:2009} and
to investigate whether the galaxy environment has an effect on the stellar distribution in the outer stellar
disc. We investigate whether the type and strength of radial $\mu$ profile breaks in the outer stellar discs
of spiral galaxies is dependent on the galaxy environment. This work builds on previous studies by using
larger and more statistically viable field and cluster samples and by reaching higher density environments.
However, it is important to note that STAGES only covers an intermediate density environment (projected
galaxy number density up to $\sim1600\,{\rm gal\,Mpc^{-3}}$, \citealt{Heiderman_etal:2009}) and not a high
density environment (e.g. the COMA cluster, $\sim 10^4\,{\rm gal\,Mpc^{-3}}$, \citealt{The_White:1986}).
We also wish to point out that in this work we use a slightly different profile classification
scheme to that used by \cite{Pohlen_Trujillo:2006} and \cite{Azzollini_etal:2008}, see Section 4.3. Therefore,
direct comparisons of the results of this work to the previous works mentioned above cannot be made.

The structure of this paper is as follows: in Section \ref{Description of the data} we give a brief
description of the STAGES dataset relevant to this work and outline our sample selection in Section
\ref{Sample selection}. In Section \ref{Profile fitting} we describe the method used to obtain our radial
surface brightness profiles from the STAGES {\em Hubble Space Telescope} ({\em HST})/Advanced Camera for
Surveys (ACS) $V$-band imaging and explain our profile classification scheme in Section \ref{Profile
classification}. We present our results in Section \ref{Results} and finally draw our conclusions in Section
\ref{Conclusions}. Throughout this paper, we adopt a cosmology of $H_0=70\,{\rm kms^{-1}Mpc^{-1}}$,
$\Omega_\Lambda=0.7$, and $\Omega_m=0.3$, and use AB magnitudes unless stated otherwise.

\section[]{Description of the data}

\label{Description of the data}

This work is entirely based on the STAGES data published by \cite{Gray_etal:2009}. STAGES is a
multiwavelength survey that covers a wide range of galaxy environments. A complex multicluster system at
$z\sim0.167$ has been the subject of $V$-band (F606W) {\em HST}/ACS imaging covering the full
$0.5^{\circ}\times0.5^{\circ}$ ($\sim5\times5\,{\rm Mpc^2}$) span of the multicluster system. The ACS imaging
is complemented by photometric redshifts and observed-/rest-frame spectral energy distributions (SEDs) from
the 17-band COMBO-17 photometric redshift survey \citep{Wolf_etal:2003}. Extensive multiwavelength
observations using {\em Spitzer}, {\em Galaxy Evolution Explorer} ({\em GALEX}), 2 degree Field (2dF),
{\em XMM-Newton}, and the Giant Metrewave Radio Telescope (GMRT) have also been carried out.
\cite{Gray_etal:2009} have performed S{\'e}rsic profile fitting using the {\sc galfit} code
\citep{Peng_etal:2002} on all {\em HST}/ACS images and conducted simulations to quantify the completeness of
the survey, all of which are publicly available\footnote{http://www.nottingham.ac.uk/astronomy/stages}.

The COMBO-17 observations used in the STAGES master catalogue were obtained with the Wide Field Imager
(WFI) at the Max Planck Gesellschaft/European Southern Observatory (ESO) 2.2-m telescope on La Silla,
Chile (see \citealt{Wolf_etal:2003} for further details). COMBO-17 used five broad-band filters
{\em UBVRI} and 12 medium-band filters covering wavelengths from $350$--$930\,{\rm nm}$ to define detailed
optical SEDs for objects with $R \leq 24$, with $R$ being the total Vega $R$-band magnitude. Generally,
photometric redshifts from COMBO-17 are accurate to $1$ per cent in $\delta{z}/(1+z)$ at $R < 21$ which
has been spectroscopically confirmed. Photo-$z$ quality degrades for progressively fainter galaxies
reaching accuracies of $2$ per cent for galaxies with $R \sim 22$ and $10$ per cent for galaxies with
$R > 24$ \citep{Wolf_etal:2004,Wolf_etal:2008}. To date, all galaxy evolution studies on the COMBO-17 data
that use photo-$z$ defined galaxy samples restrict themselves to galaxies with $R < 24$ to ensure only
reliable redshifts are used. Stellar mass estimates derived from SED fitting the 17-band photometry
are also available for COMBO-17 galaxies \citep{Borch_etal:2006,Gray_etal:2009}.

The STAGES morphological catalogue (Gray et al. in prep.) contains 5090 galaxies in STAGES with reliable
Hubble type morphologies. All galaxies with $R<23.5$ and $z_{\rm phot}<0.4$ were visually classified by
seven members of the STAGES team into the Hubble types (E, S0, Sa, Sb, Sc, Sd, Irr) and their intermediate
classes. S0s were defined to be disc galaxies with a visible bulge but no spiral arms (smooth disc). Spiral
galaxies have visible spiral arms in the disc and the classification (Sa--Sd) represents a decreasing sequence
in the bulge-to-disc ($B/D$) ratio. Weighted average estimates of the Hubble types, ignoring bars and degrees
of asymmetry were generated. In this paper, we only consider visually classified spiral galaxies (Sa, Sb, Sc,
Sd) and intermediate spiral classes are grouped to the earlier Hubble type (e.g. Sab are considered to be Sa).

\subsection[]{Sample selection}

\label{Sample selection}

\cite{Gray_etal:2009} suggest a cluster sample for STAGES defined solely from photometric redshifts. The
photo-$z$ distribution of cluster galaxies was assumed to follow a Gaussian, while the field distribution
was assumed to be consistent with the average galaxy counts $N(z, R)$ outside the cluster and to vary
smoothly with redshift and magnitude. The cluster sample is defined by a redshift interval $z_{\rm
phot}=[0.17-\Delta{z},0.17+\Delta{z}]$, where the half-width $\Delta{z}$ was allowed to vary with 
$R$-magnitude. A narrow redshift range is adopted for bright $R$-magnitudes due to the high precision of
COMBO-17 photometric redshifts, but the interval increases in width towards fainter $R$-magnitudes to
accommodate for the increase in photo-$z$ error. The completeness and contamination of the cluster sample at
all magnitude points was calculated using the counts of the smooth models \citep[see Fig. 14
from][]{Gray_etal:2009} and the photo-$z$ half-width compromised so the completeness was $>90$ per cent at
any magnitude. Contamination ($< 25$ per cent) is defined to be the fraction of field galaxies in the
cluster sample {\em at} a given magnitude (and not below it). The half-width as a function of magnitude $R$
is

\begin{equation}
\Delta{z}(R) = \sqrt{0.015^2 + 0.0096525^2(1 + 10^{0.6(R_{\rm tot} - 20.5)})}.
\end{equation}

This equation defines a photo-$z$ half-width that is limited to 0.015 at bright $R$-magnitudes but increases
as a constant multiple of the estimated photo-$z$ error at the faint end. The completeness of this selection
converges to nearly $100$ per cent for bright galaxies; see \cite{Gray_etal:2009} for further details. The
catalogue published by \cite{Gray_etal:2009} contains a number of flags ({\em combo\_flag, phot\_flag,
stages\_flag}) that allows the selection of various galaxy samples. In this study, we use a similar sample
selection to that used by \cite{Maltby_etal:2010}.

For our cluster sample we use the above cluster definition ({\em combo\_flag} $\geq 4$), we also only use
galaxies with reliable photometry (i.e. those with {\em phot\_flag} $<8$), and those defined as extended
{\em HST} sources in STAGES ({\em stages\_flag} $\geq 3$). We also limit our sample by stellar mass, cutting
at ${\rm log}\,M_*/{\rm M_{\odot}}>9$, and only select galaxies with visually classified Hubble-type
morphologies in the range Sa-Sdm using the STAGES morphological catalogue (Gray et al. in preparation). This
cluster spiral galaxy sample contains $383$ galaxies.

For our field sample we use COMBO-17 defined galaxies ({\em combo\_flag} $\geq 3$) and apply a redshift
selection that avoids the cluster. We include a lower redshift interval at $z = [0.05, 0.14]$ and an upper
redshift interval at $z = [0.22, 0.30]$, based on a similar sample selection used by \cite{Wolf_etal:2009}.
We also only use galaxies with reliable photometry (i.e. those with {\em phot\_flag} $<8$), and those
defined as extended {\em HST} sources in STAGES ({\em stages\_flag} $\geq 3$). We also limit our sample by
stellar mass, cutting at ${\rm log}\,M_*/{\rm M_{\odot}}>9$, and only select galaxies with visually classified
Hubble-type morphologies in the range Sa-Sdm using the STAGES morphological catalogue (Gray et al. in
preparation). This field spiral galaxy sample contains $318$ galaxies.

The catalogue published by \cite{Gray_etal:2009} contains two sets of derived values for properties such as
magnitude and stellar mass, one based on the photo-$z$ estimate and another assuming the galaxy is located at
the cluster redshift of $z=0.167$. This prevents the propagation of photo-$z$ errors into physical values.
Here we use the fixed redshift set of values for the cluster sample, but the original estimates for our
field comparison sample.

\begin{figure}
\includegraphics[width=0.39\textwidth]{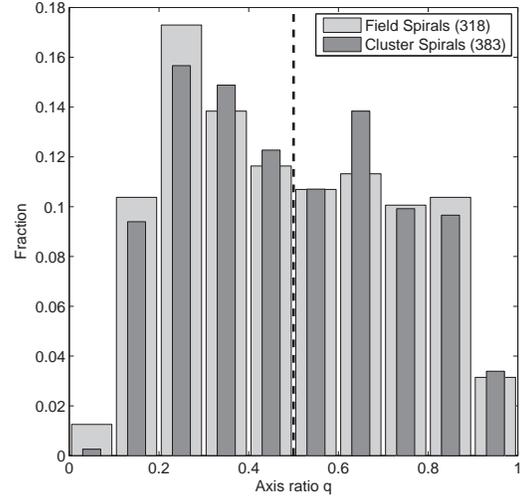}
\centering
\caption{\label{Inclination} The distribution of minor-to-major axis ratio $q$ for our field ({\em light grey})
and cluster ({\em dark grey}) spiral samples. The $q$ cut ($q > 0.5$, represented by a {\em black dashed
line}) ensures the selection of face-on to intermediately inclined spiral galaxies ($i < 60$ degrees). Errors
in $q$ are $< 3$ per cent. Respective sample sizes are shown in the legend.}
\end{figure}

\begin{figure}
\includegraphics[width=0.41\textwidth]{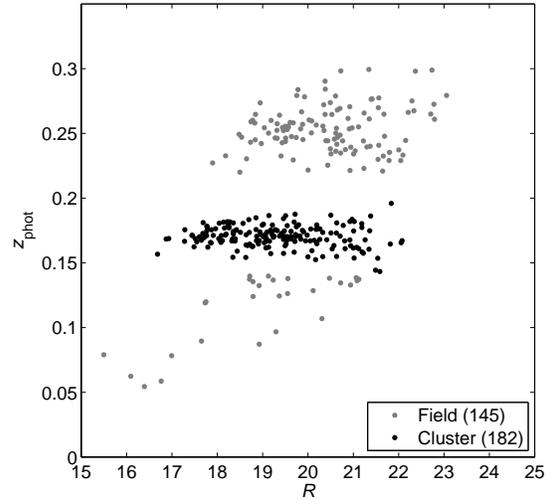}
\centering
\caption{\label{Sample} The photometric redshift estimate versus total $R$-band magnitude for the final
field ({\em grey points}) and cluster ({\em black points}) spiral galaxy samples. The field sample reaches
$R\sim23$ and the cluster sample reaches $R\sim22$. Respective sample sizes are shown in the legend.}
\end{figure}

\begin{figure}
\includegraphics[width=0.39\textwidth]{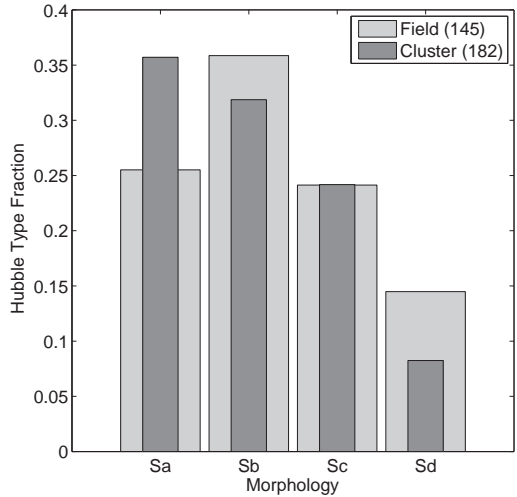}
\centering
\caption{\label{Morph mix} The distribution of Hubble-type morphologies for the final field ({\em light
grey}) and cluster ({\em dark grey}) spiral galaxy samples. Respective sample sizes are shown in the
legend. A significant excess of earlier types (Sa) is observed in the cluster environment.}
\end{figure}

The field and cluster samples are also restricted by galaxy inclination $i$ to select face-on to intermediately
inclined spiral galaxies by using the axis ratio $q$ of the galaxy as determined by the STAGES {\sc galfit}
models \citep{Gray_etal:2009}. The minor-to-major axis ratio $q$ ($q = b/a = 1-e$, where $a$ and $b$ are the
semi-major and semi-minor axes respectively and $e$ is the ellipticity) is restricted to correspond to an
inclination $i$ of less than 60 degrees ($q > 0.5$ or $e < 0.5$). This cut is necessary to avoid the influence
of dust on our surface brightness profiles and allows reliable information on features like bars, rings, and
spiral structure. The distribution of axis ratio $q$ for our field and cluster spiral samples showing the axis
ratio $q$ cut is shown in Fig.~\ref{Inclination}. This cut removes $\sim54$ per cent of field spirals and
$\sim52$ per cent of cluster spirals from our samples. The final cluster sample contains $182$ spiral galaxies
reaching down to $R\sim22$ and has a photo-$z$ range of $z_{\rm phot} = [0.143, 0.196]$. The final field sample
contains $145$ spiral galaxies reaching down to $R\sim23$ (see Fig.~\ref{Sample}). The morphological mix
(Sa, Sb, Sc, Sd) of the final field and cluster spiral samples are shown in Fig.~\ref{Morph mix}.

STAGES is $>90$ per cent complete for $R < 23.5$ \citep{Gray_etal:2009} as is the case for our field and
cluster galaxy samples (see Fig.~\ref{Sample}). However, \cite{Wolf_etal:2009} estimate that at stellar masses
below ${\rm log}\,M_*/{\rm M_\odot} < 9.5$, the field sample could have an additional $20$ per cent incompleteness
based on previous COMBO-17 experience. Therefore, our field sample is essentially $>70$ per cent complete.
We expect the completeness of the cluster sample to be $>90$ per cent and the contamination of the cluster
sample by the field to be $<25$ per cent, based on the $R$-magnitude the respective samples reach
\citep[see Fig. 14 from][]{Gray_etal:2009}. Selecting spiral galaxies from the STAGES morphological catalogue
(Gray et al. in prep) introduces no further incompleteness to the galaxy samples \citep{Maltby_etal:2010}.
The properties of the final field and cluster spiral samples are shown in Table.~\ref{Sample tbl}.

\begin{table}
\begin{minipage}{80mm}
\centering
\caption{\label{Sample tbl}{Properties of the field and cluster spiral samples.}}
\begin{tabular}{lccc}
\hline
\hline
Property		& Field			& Cluster		\\
\hline
$N_{\rm gal}$		& $145$			& $182$			\\
${\rm Completeness}$	& $>70$ per cent	& $>90$ per cent	\\
${\rm Contamination}$	& $-$			& $<25$ per cent	\\
$R_{\rm mean}$		& $20.14$		& $19.32$		\\
$M_{B\rm (min)}$	& $-15.1$		& $-16.1$		\\
$M_{B\rm (max)}$	& $-21.5$		& $-21.8$		\\
$z_{\rm phot,mean}$		& $0.227$		& $0.171$		\\
$z_{\rm phot,min}$	& $0.055$		& $0.143$		\\
$z_{\rm phot,max}$	& $0.299$		& $0.196$		\\
\hline
\end{tabular}
\end{minipage}
\end{table}

\section[]{Profile fitting}

\label{Profile fitting}

We use the IRAF task {\em ellipse} (STSDAS package - version 2.12.2) in order to obtain azimuthally averaged
radial surface brightness profiles from the STAGES {\em HST}/ACS $V$-band imaging. Profile fitting is
carried out for each spiral galaxy in our field and cluster samples. The ACS images used include the sky
background and the necessary sky subtraction is performed after the profile fitting (see Section \ref{Sky
subtraction}). For further details on the fitting method used by {\em ellipse} see \cite{Jedrzejewski:1987}.

\begin{figure*} 
\includegraphics[width=0.75\textwidth]{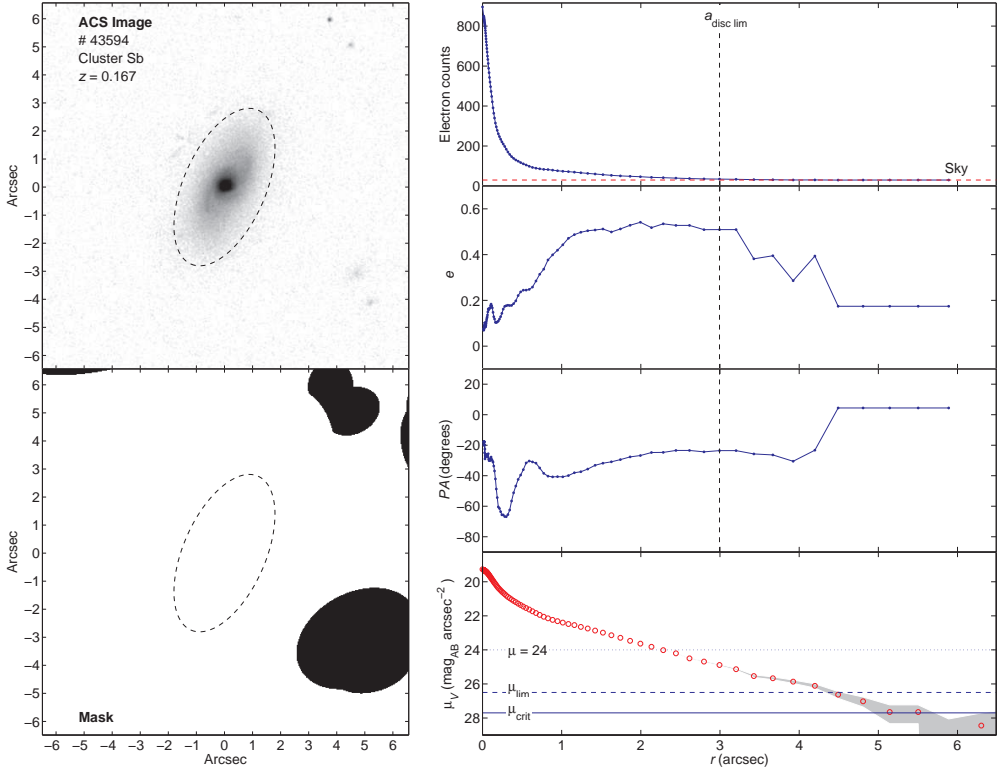}
\caption{\label{Type I fit} Example profile fit. {\em Left column:} ACS image ({\em top}) and bad pixel mask
({\em bottom}) showing the visually determined stellar disc limit $a_{\rm disc\,lim}$ ({\em black dashed
line}). The ACS image is shown with a logarithmic greyscale. {\em Right column, top panel:} The azimuthally
averaged radial electron count $n_{\rm e}$ profile from the free-parameter fit showing the {\sc galapagos}
sky background ({\em red dashed line}) and the
stellar disc limit $a_{\rm disc\,lim}$ ({\em black dashed line}). {\em Right column, second and third
panels:} The ellipticity $e$, and position angle ${\it PA}$ radial profiles from the free-fit showing the
position of the stellar disc limit $a_{\rm disc\,lim}$ ({\em black dashed line}) used to obtain the $e$ and
${\it PA}$ of the outer stellar disc. ${\it PA}$ is measured east of north. {\em Right column, bottom panel:}
The final azimuthally averaged radial surface brightness profile from the fixed-parameter fit. The error in
the surface brightness ({\em grey shaded area}) is due to over- and undersubtracting the sky background by
$\pm1\sigma$. The limiting surface brightness $\mu_{\rm lim}$ ({\em blue dashed line}) and critical surface
brightness $\mu_{\rm crit}$({\em blue solid line}) represent the limit to which we trust profile breaks and
the surface brightness profile respectively.}
\end{figure*}

We run {\em ellipse} using bad pixel masks that remove flagged pixels from the isophotal fit. This is
necessary in order to remove sources of contamination such as background/companion galaxies and foreground
stars (everything not associated with the galaxy itself), see Fig.~\ref{Type I fit} for
an example. \cite{Gray_etal:2009} generated these bad pixel masks for each STAGES galaxy using the data
pipeline {\sc galapagos} (Galaxy Analysis over Large Areas: Parameter Assessment by {\sc galfit}ting Objects
from {\sc SExtractor}; Barden et al., in prep). {\sc galapagos} performs an extraction of source galaxies
from the STAGES {\em HST}/ACS $V$-band imaging and uses the {\sc galfit} code \citep{Peng_etal:2002} to fit
\cite{Sersic:1968} radial surface brightness models to each galaxy image. The bad pixel masks are generated
for each STAGES galaxy in order to remove sources of contamination from these surface brightness models.
However, occasionally in crowded regions {\sc galfit} performs multiobject fitting and therefore the
companion galaxies in these cases are not removed by the bad pixel mask as they are too close to the subject
galaxy. In these cases ($27$ field and $33$ cluster galaxies, $\sim18$ per cent) the companion galaxies are
removed from the ACS image by subtraction of their {\sc galfit} surface brightness model. The residuals of
the companion galaxies are not expected to have any significant affect on the azimuthally averaged radial
surface brightness profile for the subject galaxy. Isophotal fitting is only attempted when the fraction of
flagged (masked) data points in the isophote is less than $0.5$ and fitting is terminated if this condition
is broken.

For all our {\em ellipse} isophotal fits the galaxy centre is fixed (all isophotes have a common centre)
using the centre of the galaxy determined from the {\sc galfit} S\'{e}rsic model \citep{Gray_etal:2009}.
We also use logarithmic radial sampling with steps of $0.03$ dex ($0.07$ geometric sampling in {\em ellipse}
terminology) and start from an initial semi-major axis of $10$ pixels.

For each galaxy in our sample we fit two different sets of ellipses to the galaxy ACS image. The first is a
free-parameter fit (fixed centre, free ellipticity $e$ and position angle ${\it PA}$) and tends to follow
morphological features like bars and spiral arms, and therefore is not ideal for the characterisation of
the underlying outer stellar disc studied in this paper. Therefore, we use a fixed-parameter fit (fixed
centre, $e$, and ${\it PA}$) in order to produce our final surface brightness profiles.

The initial free-parameter fit (fixed centre, free $e$ and ${\it PA}$) is used to determine the ellipticity
and position angle of the outer disc component. For each sample galaxy the ACS image was inspected with the
contrast adjusted in order to get an estimate of the semi-major axis for the end of the stellar disc $a_{\rm
disc\,lim}$ (where the galaxy surface brightness enters the background
noise). This `stellar disc limit' $a_{\rm disc\,lim}$ corresponds to the outer stellar disc region and is used to
obtain the ellipticity and position angle of the outer stellar disc, $e_{\rm disc\,lim}$ and ${\it PA}_{\rm
disc\,lim}$ respectively, from the {\em ellipse} free-fit $e$ and ${\it PA}$ radial profiles (see
Fig.~\ref{Type I fit}). Fig.~\ref{Type I fit} shows the fitting procedure for an example galaxy.

In some cases the fit failed at the stellar disc limit $a_{\rm disc\,lim}$ due to {\em ellipse} error
limits in the radial intensity gradient. However, adjusting this threshold allowed these fits to be forced
out to $a_{\rm disc\, lim}$. In order to ensure that the ellipticity $e_{\rm disc\,lim}$ and position angle
${\it PA}_{\rm disc\,lim}$ of the outer stellar disc are reliable, we compared the values obtained from the
{\em ellipse} fit to a manual measurement of $e_{\rm disc\,lim}$ and ${\it PA}_{\rm disc\,lim}$ measured
directly off the ACS image during the visual inspection. The {\em ellipse} values agreed with the visually
measured values in all cases.

However, in some cases the {\em ellipse} radial $e$ and ${\it PA}$ profiles could be unstable at the stellar
disc limit $a_{\rm disc\,lim}$. In these cases, $7$ field ($\sim 5$ per cent) and $13$ cluster galaxies
($\sim 7$ per cent), we use the last reliable $e$ and ${\it PA}$ obtained from the {\em ellipse} profiles as
our estimate of $e$ and ${\it PA}$ for the outer stellar disc.

In some other cases the free-fit failed completely at $a_{\rm disc\,lim}$. Subsequent inspection of the ACS
image sometimes revealed clear structure at $a_{\rm disc\,lim}$. In these cases, $27$ field ($\sim 19$ per
cent) and $24$ cluster galaxies ($\sim 13$ per cent), we use the estimate of $e_{\rm disc\,lim}$ and ${\it
PA_{\rm disc\,lim}}$ measured directly off the ACS image as our estimate of $e$ and ${\it PA}$ for the outer
stellar disc. Unfortunately, in a few cases the ACS image was too noisy for a reliable manual measurement of
$e_{\rm disc\,lim}$ and ${\it PA_{\rm disc\,lim}}$. In these cases, $7$ field ($\sim 5$ per cent) and $4$
cluster galaxies ($\sim 2$ per cent), we use the value of the $e$ and ${\it PA}$ determined by {\sc galfit}
in the STAGES S\'{e}rsic models \citep{Gray_etal:2009} as our estimate of $e$ and ${\it PA}$ for the outer
stellar disc.

A fixed-parameter fit (fixed centre, $e$, and ${\it PA}$) using the $e$ and ${\it PA}$ of the outer stellar
disc obtained from the free-fit is then used to produce our final surface brightness profiles (see
Fig.~\ref{Type I fit}). The fixed fits are forced to extend out well into the sky
background (semi-major axis of $600$ pixels). During the fixed-fitting process we also perform 4 iterations
of a $3-\sigma$ rejection applied to deviant points below and above the average to smooth some of the bumps
in the surface brightness profiles that are due to non-axisymmetric features, i.e. not part of the disc (e.g.
star-forming regions and supernovae). The final profile fits then undergo a sky subtraction and surface
brightness calibration in order to produce the final surface brightness profiles which will be discussed in
the following sections.


\subsection[]{Photometric calibration}

\label{Photometric calibration}

The STAGES {\em HST}/ACS $V$-band images have pixel values in units of electrons $n_{\rm e}$. These can be
converted to AB magnitudes per pixel using the following expression,

\begin{equation}
m_{\rm AB} = -2.5{\rm log}_{10} \left ( \frac {{\it PHOTFLAM} \times n_{\rm e}} {\it EXPTIME} \right ) +
zpt_{\rm AB},
\end{equation}

where the AB zeropoint

\begin{equation}
zpt_{\rm AB} = {\it PHOTZPT} - 5{\rm log}_{10}({\it PHOTPLAM}) + 18.6921.
\end{equation}

The following FITS header keywords are used (ACS Datahandbook v5.0\footnote{Pavlovsky, C., et al. 2004, "ACS
Data Handbook", Version 3.0, (Baltimore: STScI)}),

\begin{math}
											\\
{\it PHOTFLAM} = 7.766405\times10^{-20}\,\rm erg\,cm^{-2}\,\AA^{-1}\,electron^{-1},	\\
{\it PHOTPLAM} = 5.919369\times10^3 \AA,						\\
{\rm and}\:{\it PHOTZPT} = -21.10.							\\
\end{math}

The effective exposure time ${\it EXPTIME}$ (after ACS subpixel drizzling) varies for each sample galaxy
and is between $700$--$770\, {\rm seconds}$. The AB magnitude for each pixel $m_{\rm AB}$ is then converted
into a surface brightness $\mu$ using the ACS pixel scale ($0.03\,''$),

\begin{equation}
\mu = m_{\rm AB} + 2.5{\rm log}_{10}({\it pixel\:scale^2}).
\end{equation}

This gives a surface brightness $\mu$ per pixel in $\rm mag_{AB}\,{\rm arcsec}^{-2}$ in the $V$-band.

Several corrections are also required in order to correct the surface brightness $\mu$ for galactic
extinction, individual galaxy inclination $i$, and surface brightness dimming.

\begin{enumerate}
\item {\em Galactic extinction:} The STAGES field is affected by reddening due to substantial foreground dust
\citep[galactic extinction, see][]{Gray_etal:2009} and an  extinction correction $A_{\rm extinction}$ of
$-0.18$ magnitudes in the $V$-band is required.

\item {\em Galaxy inclination:} The surface brightness $\mu$ of a spiral galaxy increases with its inclination
$i$ to the line of sight. We correct for galaxy inclination using the surface brightness correction
$A_i = -2.5{\rm log}_{10}(1-e)$, where $e$ is the ellipticity of the outer isophote \citep{Freeman:1970}.
We use the ellipticity determined for the fixed-fit $e_{{\rm disc\,lim}}$ (outer isophote) to correct for
galaxy inclination $i$. We do not attempt to correct the surface brightness for internal extinction (dust
effects). However, since the galaxies studied here are all reasonably face-on ($\langle{q}\rangle = 0.7$,
$\langle{i}\rangle = 45^{\circ}$), and as we are dealing with
the outer parts of the galactic disc, dust is expected to have very little effect on the surface brightness
correction and no effect on the presence or strength of truncations/anti-truncations in the stellar disc. 

\item {\em Surface brightness dimming:} We correct the surface brightness $\mu$ profiles of our field galaxies
so they are at the redshift of the cluster ($z_{\rm cl} = 0.167$). This allows a fair comparison between
galaxies across the redshift range of our sample. All cluster member galaxies are assumed to be at the
cluster spectroscopic redshift. Surface brightness $\mu\propto(1+z)^{-4}$ for bolometric luminosity; however
as we use the $V$-band (F606W) filter (effective filter width dependent on $z$) we loose a $(1+z)$ term from
this relation and $\mu\propto(1+z)^{-3}$. Therefore, correcting the $\mu$ of our field galaxies requires
multiplying the galaxy flux (in electrons $n_{\rm e}$) by $(1+z_{\rm gal})^3/(1+z_{\rm cl})^3$, where
$z_{\rm gal}$ is the photometric redshift of the field galaxy from COMBO-17. This leads to a surface
brightness correction $A_z$ for our field galaxies where

\begin{equation}
A_z = -2.5{\rm log}_{10} \left ( \frac {(1+z_{\rm gal})^3}{(1+z_{\rm cl})^3} \right ).
\end{equation}

Therefore, the final surface brightness $\mu_{\rm corrected}$ corrected for galactic extinction, galaxy
inclination $i$, and surface brightness dimming is given by

\begin{equation}
\mu_{\rm corrected} = \mu_{\rm measured} + A_{\rm extinction} + A_i + A_z.
\end{equation}

\end{enumerate}

\subsection[]{Sky subtraction}

\label{Sky subtraction}

During the {\sc galfit} S\'{e}rsic model fitting performed by the {\sc galapagos} pipeline
\citep{Gray_etal:2009}, the sky level is calculated individually for each source galaxy by evaluating a flux
growth curve and using the full science frame. In this paper, for each sample galaxy we use the sky level
determined by {\sc galapagos} for our sky subtraction. Fig.~\ref{Type I fit} includes the free-fit radial
intensity profile (in electron counts $n_e$) for an example galaxy showing the sky background determined by
{\sc galapagos}.

The error in the {\sc galapagos} sky level was estimated using observed `dark' patches of sky located in
some of the STAGES ACS tiles \citep[tiles $01$, $10$, $31$, and $60$, see][]{Gray_etal:2009}. Each patch
(measuring $100 \times 100$ pixels) was visually inspected to ensure it was clear of any $V$-band sources.
The distribution of electron counts $n_{\rm e}$ (pixel values) in each dark patch is Gaussian in all cases.
For each ACS tile we obtain the mean pixel value $\langle{n_{\rm e}}\rangle$ in each dark patch of sky and
then the standard deviation in the mean pixel values $\sigma_{\langle{n_{\rm e}}\rangle}$, see
Table~\ref{sky error tbl}. The distribution of mean pixel values $\langle{n_{\rm e}}\rangle$ is approximately
Gaussian for each tile. The final $\pm1\sigma$ sky error is the mean $\sigma_{\langle{n_{\rm e}}\rangle}$
from the 4 selected ACS tiles and is $\sim\pm0.18$ electron counts, see Table~\ref{sky error tbl}.

\begin{table}
\begin{minipage}{80mm}
\centering
\caption{\label{sky error tbl}{Determination of the $\pm1\sigma$ sky error from observed dark patches of sky
in the STAGES ACS $V$-band imaging.}}
\begin{tabular}{ccc}
\hline
\hline
STAGES tile	& $N_{\rm Dark\:patches}$			& $\sigma_{\langle{n_{\rm e}}\rangle}$ (electron counts)	\\
\hline
$01$		& $24$						& $0.1191$				\\
$10$		& $28$						& $0.1314$				\\
$31$		& $23$						& $0.3403$				\\
$60$		& $30$						& $0.1110$				\\
\hline
\multicolumn{2}{c}{$\pm1\sigma$ sky error =  Mean $\sigma_{\langle{n_{\rm e}}\rangle}$}	& $0.1754$	\\
\hline
\end{tabular}
\end{minipage}
\end{table}

The sky subtraction error in the surface brightness profiles due to the error in the {\sc galapagos} sky
background dominates over the individual errors produced by {\em ellipse} in the fitting process. At $\mu < 25
\,\rm mag\,arcsec^{-2}$ the fit error dominates over the error in the sky subtraction but has a negligible effect on
the surface brightness profile. However, at $\mu > 25\,\rm mag\,arcsec^{-2}$ the sky subtraction error dominates the
error in the surface brightness profile. The sky subtraction error can have a significant effect on the
surface brightness profile of the spiral galaxies, especially in the outer regions where the surface
brightness $\mu_V$ approaches that of the sky background. However, for any particular galaxy the global sky
subtraction error is approximately constant across the length of the surface brightness profile. Therefore we can
specify the error in our surface brightness profiles by generating profiles for when the sky background is over- and
undersubtracted by $\pm1\sigma$ (see Fig.~\ref{Type I fit}). 

The $\pm1\sigma$ error in the sky background corresponds to a critical surface brightness limit $\mu_{\rm
crit}$ below which the sky subtracted radial surface brightness profile of a galaxy becomes unreliable. This
critical surface brightness $\mu_{\rm crit}$ is approximately $27.7\,{\rm mag\,arcsec}^{-2}$. We also define
a limiting surface brightness $\mu_{\rm lim}$, corresponding to a $\pm3\sigma$ sky error, below which
identifying profile breaks becomes unreliable. The limiting surface brightness $\mu_{\rm lim}$ is
approximately $26.5\,{\rm mag\,arcsec^2}$.

\section[]{Profile classification}

\label{Profile classification}

\subsection{Profile inspection}

\label{Profile inspection}

The azimuthally averaged radial surface brightness profile for each spiral galaxy in our field and cluster
samples was visually inspected by three independent assessors (DTM, AAS, MEG) in order to identify potential
profile breaks (inflection points in the exponential region of the $\mu $ profile). Three possible
cases were considered: no break or simple exponential profile, a single broken exponential, and cases with
two profile breaks. In each case, break identification relates to the outer disc component of the galaxy
radial surface brightness $\mu$ profile and does not consider the inner varying bulge component. 

Our break identification is based solely on the surface brightness profiles and without direct inspection of
the ACS images. We chose not to relate the profile breaks to visually identified structural features because
we wanted a break classification method that treated all galaxies equally, in a self-consistent manner, and
avoided the prejudice that image inspection could introduce. Additionally, the aims of this work are to
explore the effect of the environment on the structure of the outer regions of the galactic disc, regardless
of the origins of any identified structural features.

The radial limits of exponential regions either side of the break radius $r_{\rm brk}$ were also identified
allowing for some bumps and wiggles due to spiral substructure (e.g. spiral arms and star forming regions).
For the inner exponential disc, the inner boundary is chosen manually to avoid the region dominated by the
bulge component. For the outer exponential disc, the outer boundary is generally taken to be where the surface
brightness profile reaches the critical surface brightness ($1\sigma$ above the sky background) but may
be at higher $\mu$ depending on the nature of the profile. A small
manually selected transition region (non-exponential) is allowed between exponential regions either side of
the break. The break radius $r_{\rm brk}$ is defined as the mean radius for the radial limits of this
transition region. 

Due to the subjective nature of some galaxy profile classifications, the number of galaxies with either no,
one, or two breaks varied subtly between the different assessors. To account for this, in what follows we
perform parallel analysis on the breaks identified by each assessor and compare the final results.

\subsection{Defining the Outer Stellar Disc}

\label{Defining the Outer Stellar Disc}

\begin{figure*}
\includegraphics[width=1\textwidth]{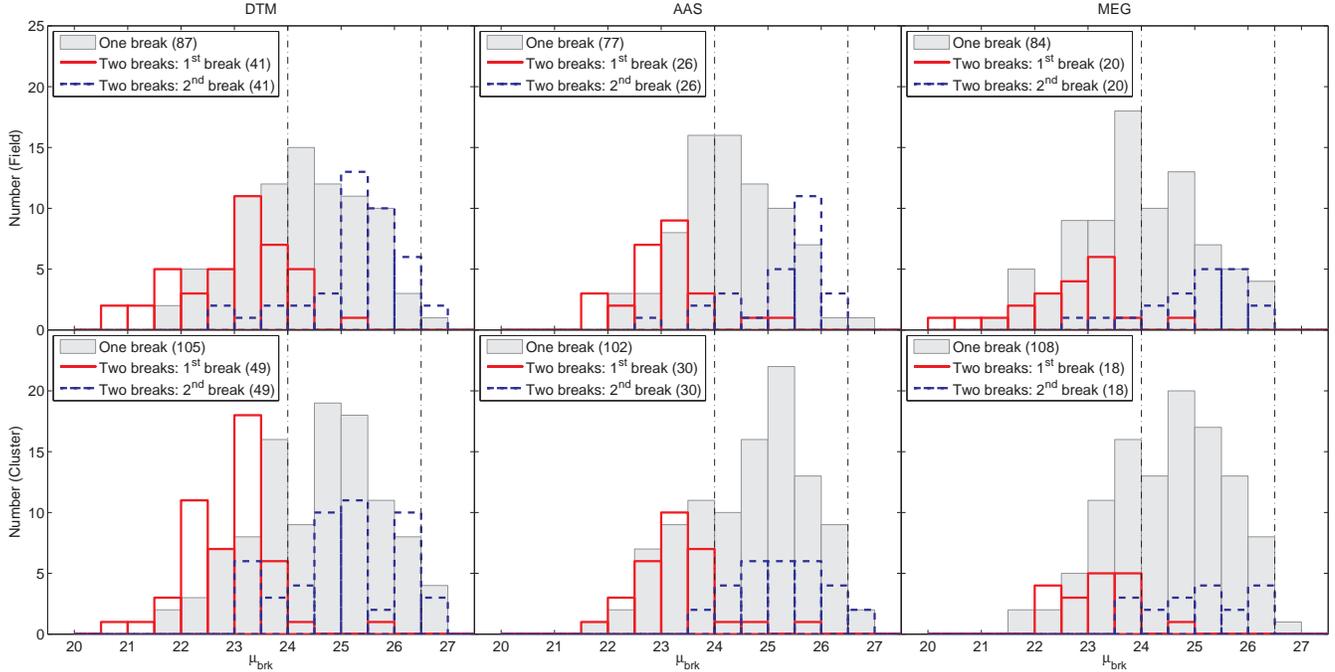}
\caption{\label{Break surface brightness} The distribution of break surface brightness $\mu_{\rm brk}$. The
surface brightness at the break radius $\mu_{\rm brk}$ for field ({\em top row}) and cluster ({\em bottom
row}) galaxies as determined by DTM ({\em left column}), AAS ({\em centre column}), and MEG ({\em right
column}). The distributions show galaxies with one break ({\em grey shaded area}), and both the inner ({\em
red line}) and outer break ({\em blue dashed line}) of galaxies with two breaks. For the case of two breaks
the separation between the inner and outer break occurs at a surface brightness of $\sim24\,{\rm
mag\,arcsec}^{-2}$. The position of the break surface brightness cut $24 < \mu_{\rm brk} < 26.5\, {\rm
mag\,arcsec}^{-2}$ used to create our final break samples is shown ({\em black dashed lines}). Respective
sample sizes are shown in the legends. Contamination of the cluster sample by the field is $< 25$ per cent.}
\end{figure*}

In this study, we wish to consider the effect of the galaxy environment on the outer regions of the stellar
disc for spiral galaxies. Therefore, we only wish to consider broken exponentials in the outer disc region
of the $\mu$ profile. Any effect of the environment should be stronger in these faint, fragile, outer
regions. Fig.~\ref{Break surface brightness} shows the surface brightness at the break radius $\mu_{\rm brk}$
for galaxies with one, and two breaks. In the case of one break, for each assessor there appears to be a
slight difference in the distribution of $\mu_{\rm brk}$ between the field and cluster samples with cluster
breaks occurring at fainter surface brightnesses. Kolmogorov-Smirnov (K--S) tests between the respective
field and cluster samples show that this environmental difference is driven by Type $\rm III$ (up-bending
break, antitruncated) profiles with a significance at the $2$--$3\sigma$ level. We offer no interpretation of
this result due to its uncertain nature. In the case of two inflection points, we consider the inner and
outer breaks separately and the separation between the inner and outer break occurs at a surface brightness $\mu\sim24\,{\rm
mag\,arcsec}^{-2}$ (see Fig.~\ref{Break surface brightness}). The same conclusion is reached for the break
samples generated by each assessor. Inner and outer breaks may have different physical origins. Therefore,
in order to ensure we are comparing intrinsically similar $\mu$ breaks in the stellar disc, we limit the
breaks analysed to those with $\mu_{\rm brk} > 24 \,\rm mag\,arcsec^{-2}$ (outer disc breaks) and to the
outermost break if two inflection points are present in this range. There may be break features at surface
brightnesses $\mu < 24\,{\rm mag\,arcsec}^{-2}$, and these may be related to the galaxy environment. However, we focus on breaks in the outer disc as these are more likely to be susceptible to environmental
effects. We acknowledge that in doing this there is some potential for missed environmental effects in the inner
disc, therefore we do include some additional tests using the inner/initial break, see Section~\ref{Inner/initial
disc breaks}.

In Section \ref{Sky subtraction} we defined a limiting surface brightness $\mu_{\rm lim}$ ($\pm3\sigma$ sky
error) of $26.5\,{\rm mag\,arcsec}^{-2}$. In our analysis we do not trust breaks at $\mu$ fainter than the
$\mu_{\rm lim}$ level as this is where identification of breaks becomes unreliable due to the deviation of
the profiles generated by over- and undersubtracting the sky by $\pm1\sigma$ (see Fig.~\ref{Type I fit}). We
therefore also restrict our analysis to profile breaks that have $\mu_{\rm brk} < 26.5\,{\rm
mag\,arcsec}^{-2}$.

\subsection{Profile Types}

\label{Profile Types}

\begin{figure*}
\includegraphics[width=0.315\textwidth]{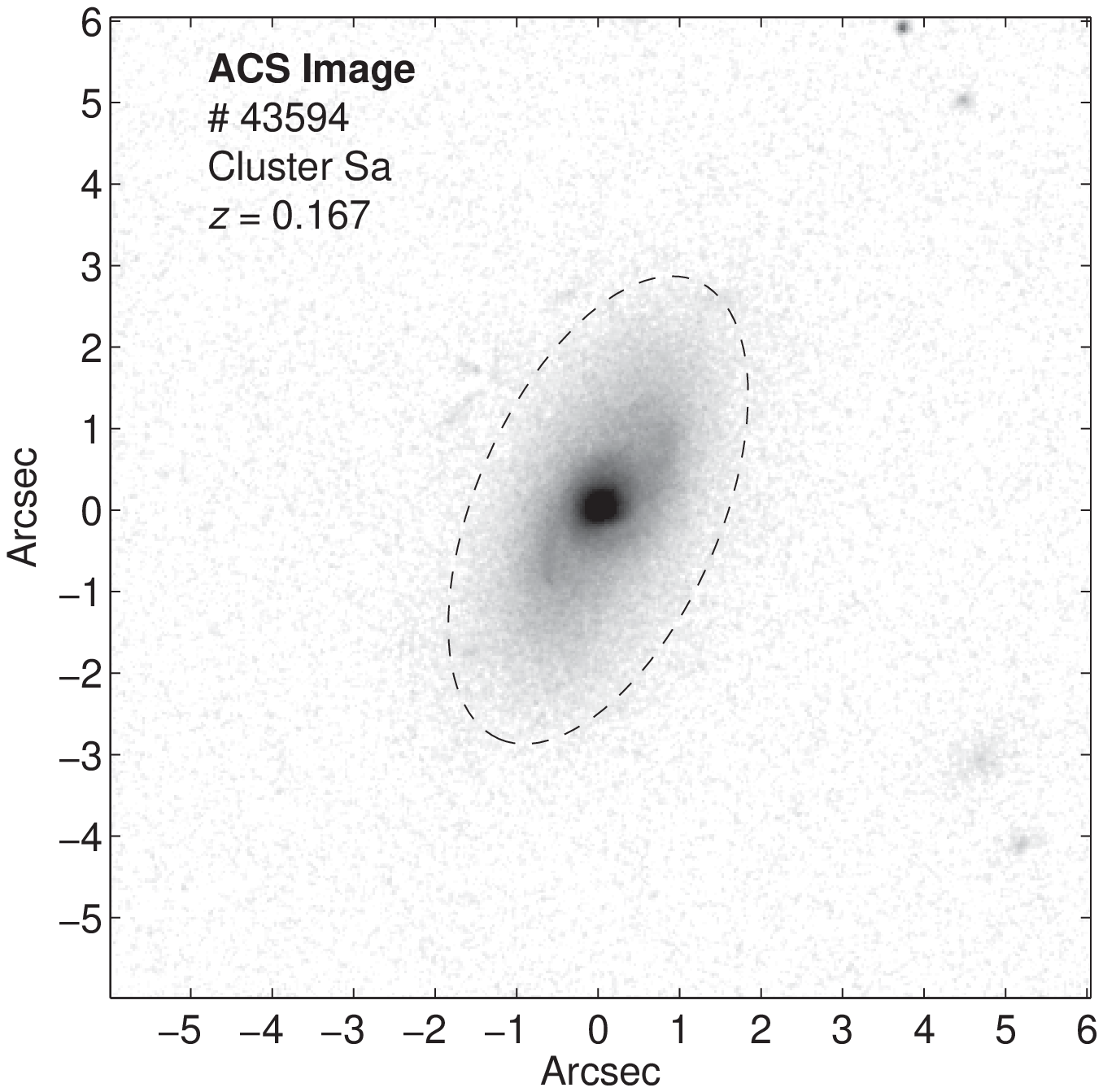}
\includegraphics[width=0.4\textwidth]{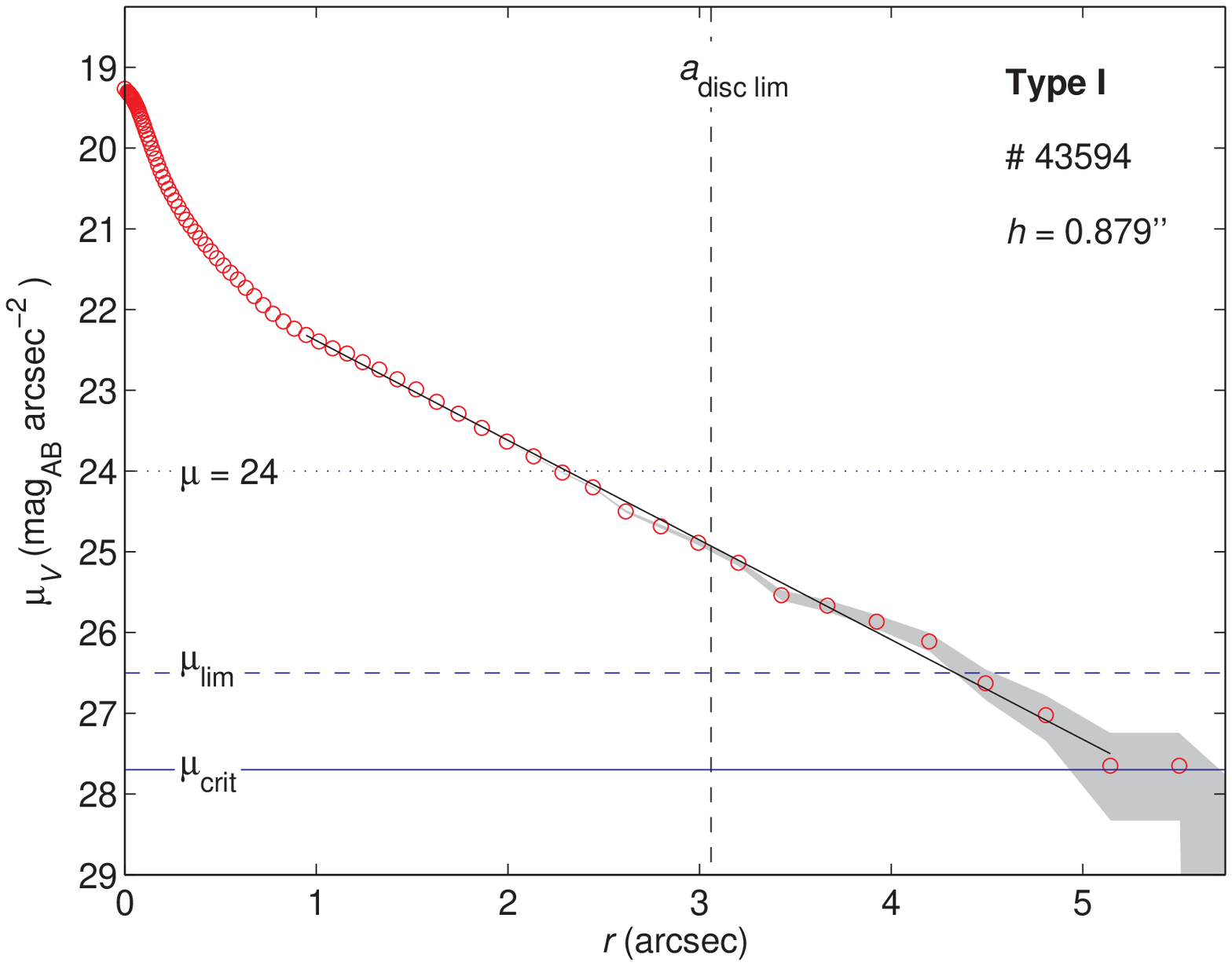}	\\
\includegraphics[width=0.315\textwidth]{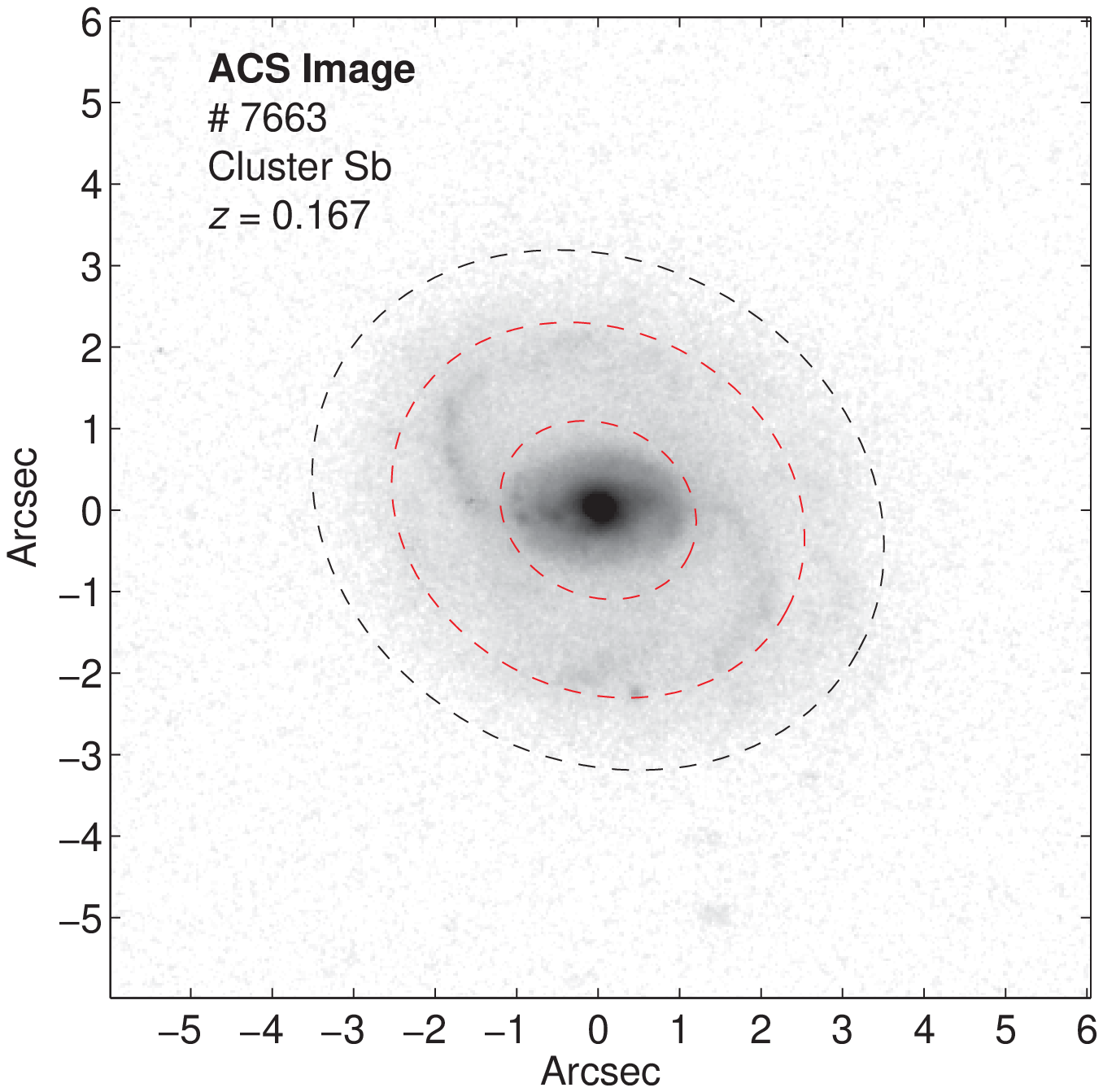}
\includegraphics[width=0.40\textwidth]{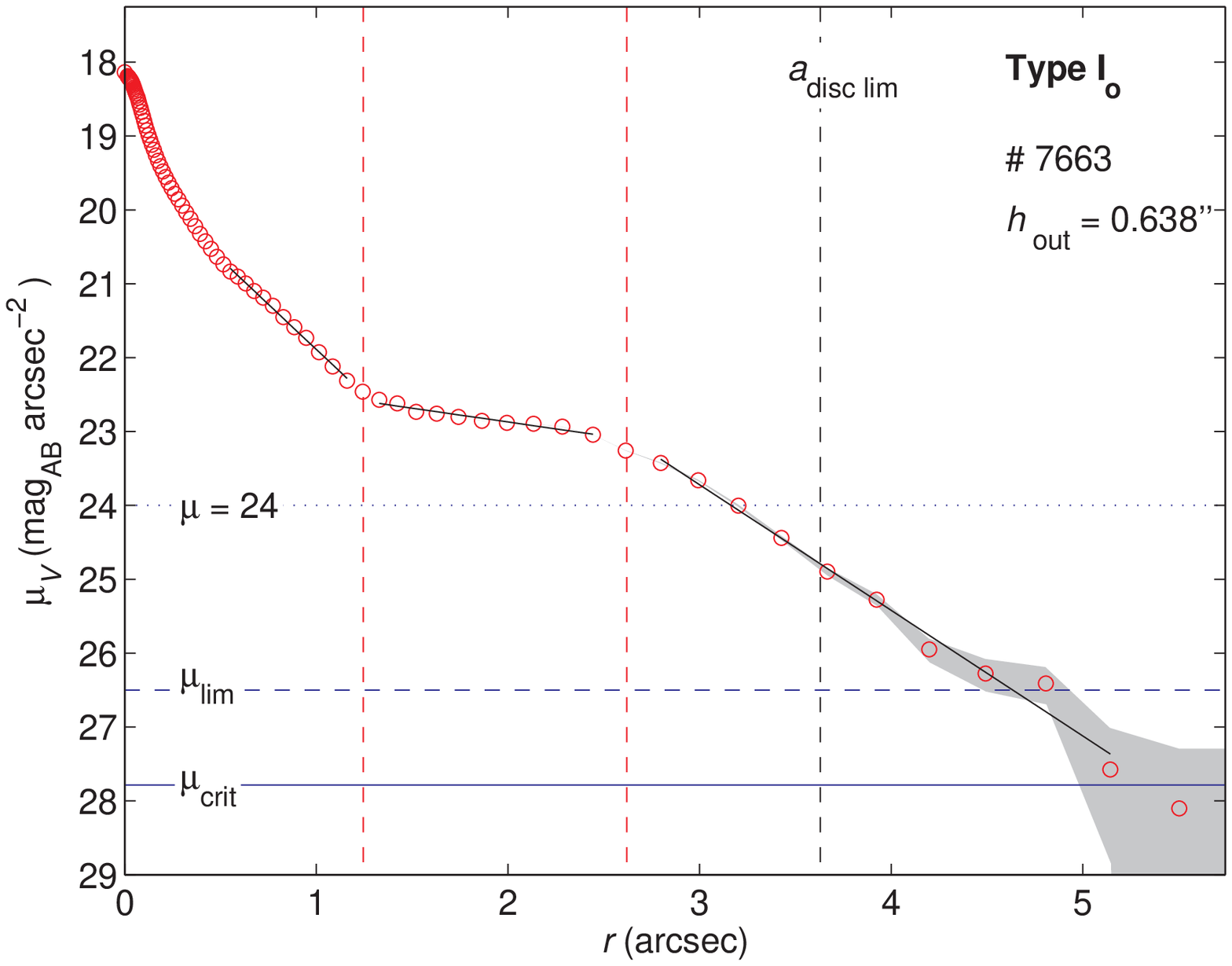}	\\
\includegraphics[width=0.315\textwidth]{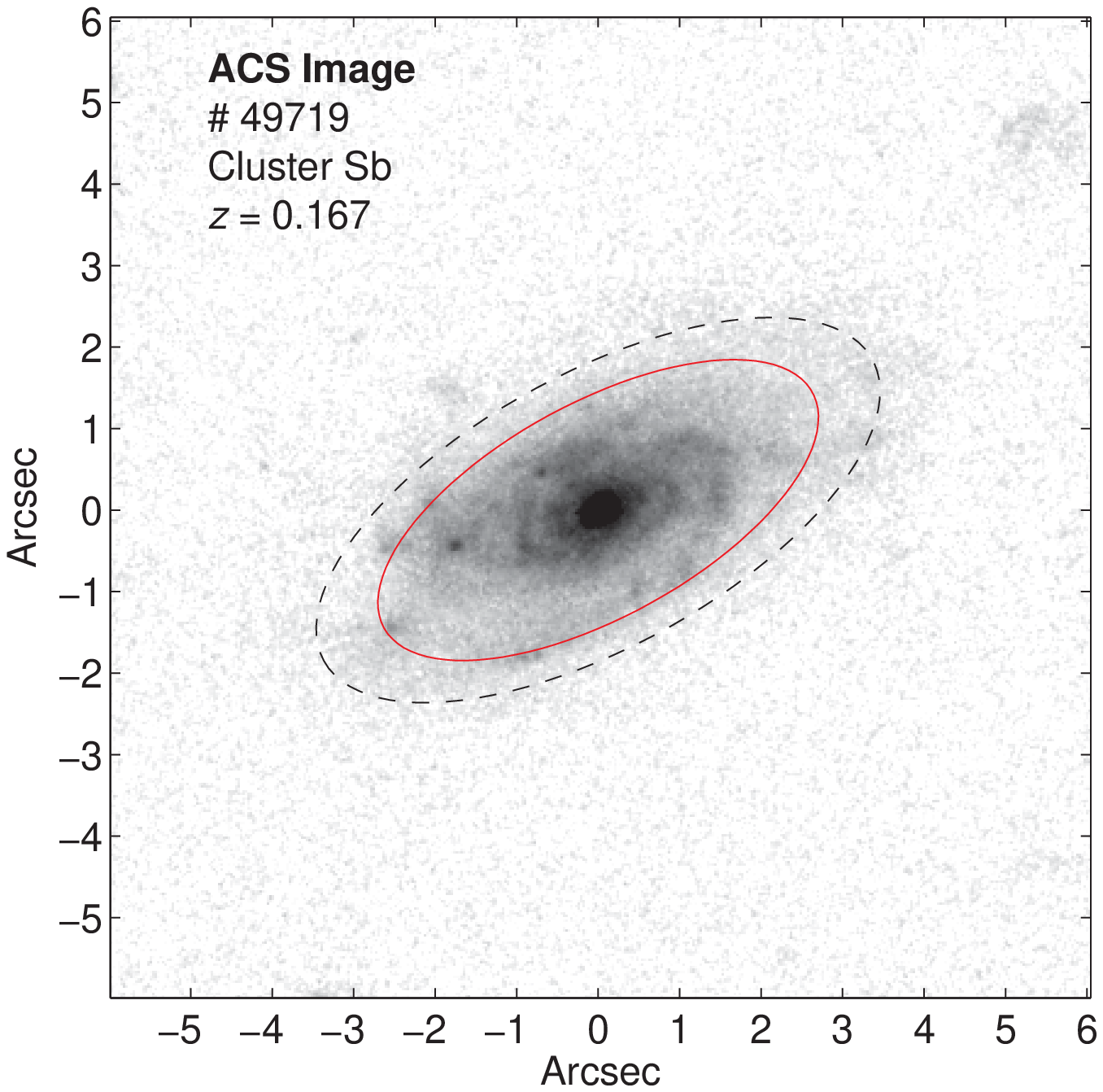}
\includegraphics[width=0.40\textwidth]{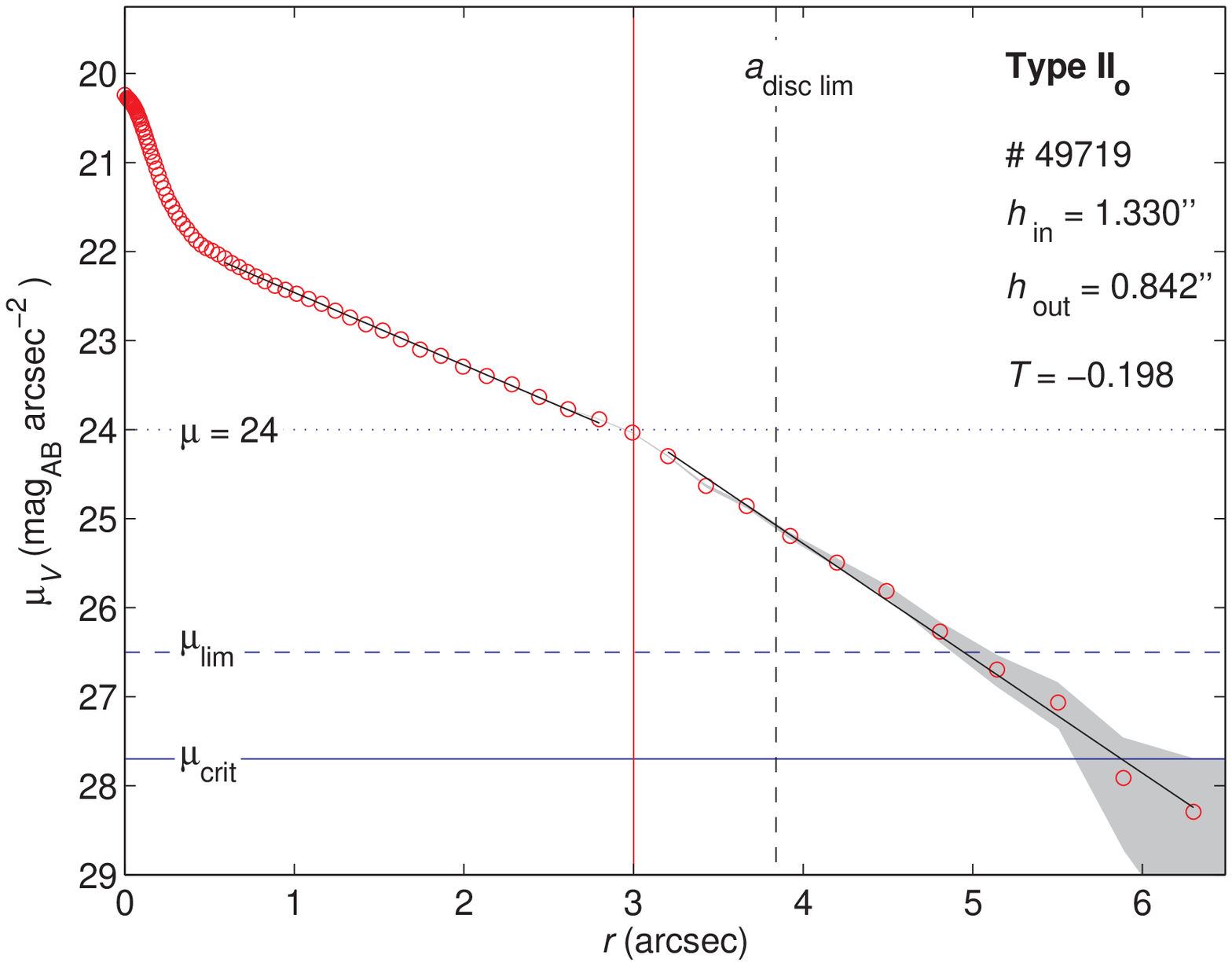}	\\
\includegraphics[width=0.315\textwidth]{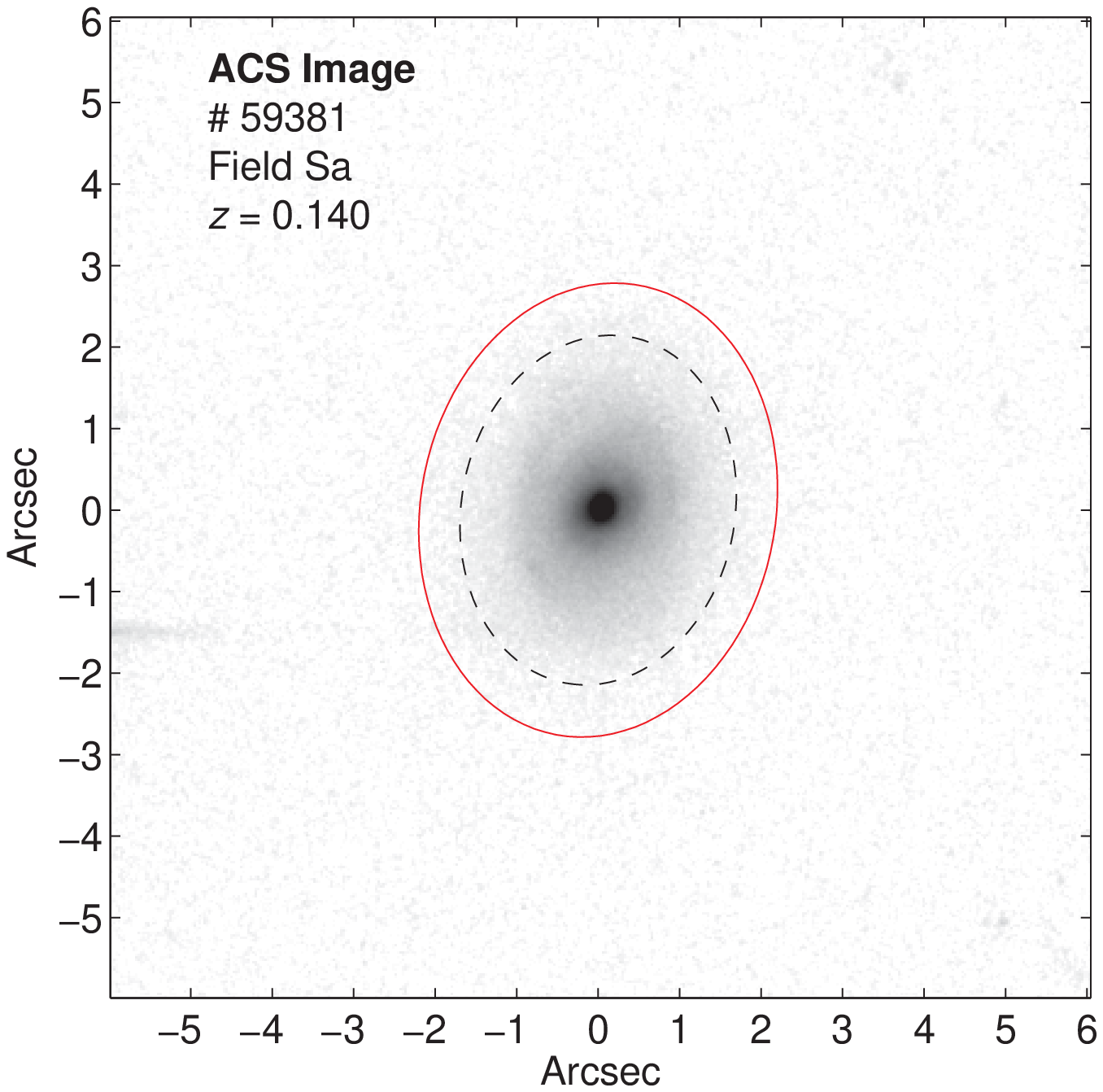}
\includegraphics[width=0.40\textwidth]{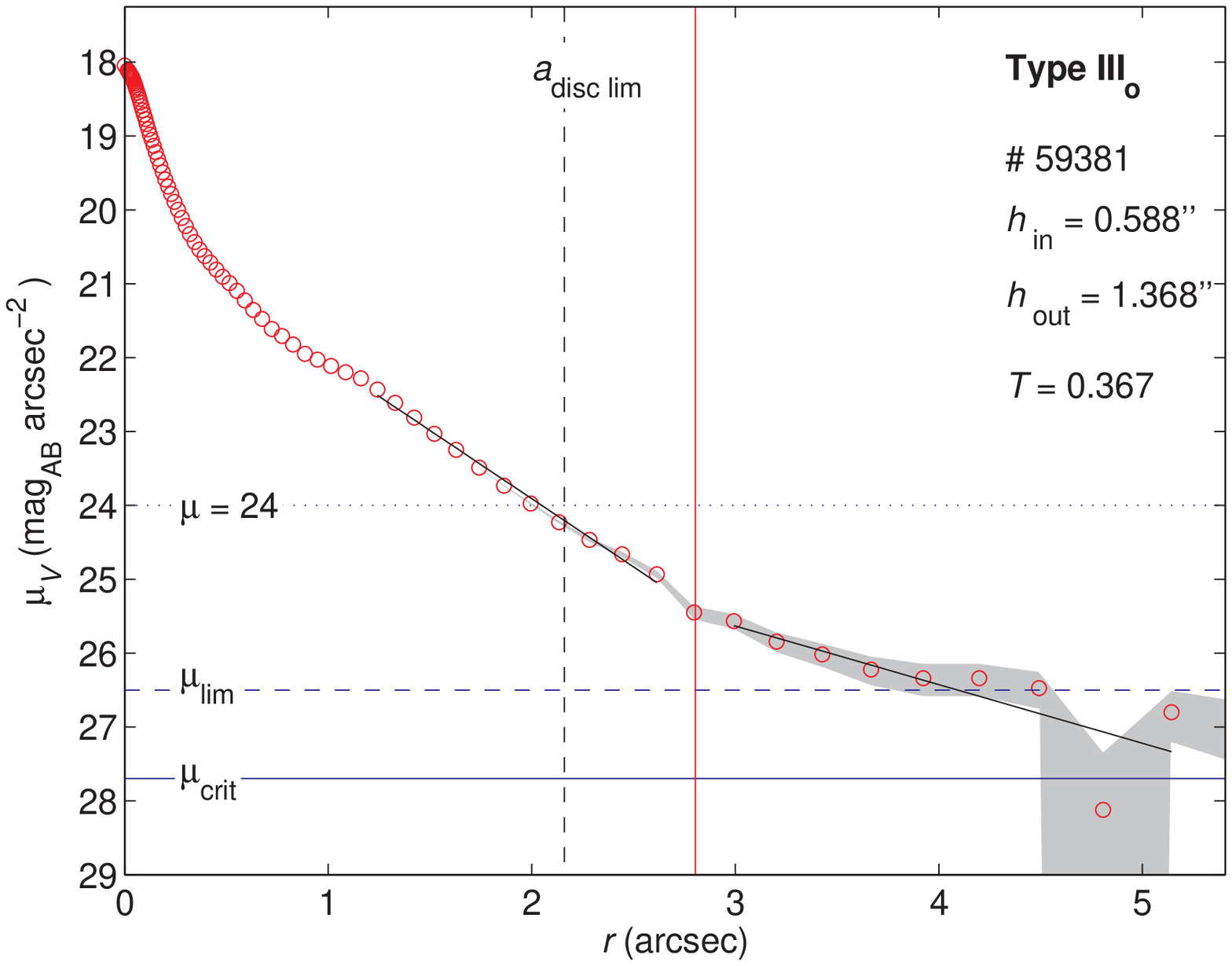}	\\
\caption{\label{Profile types} Examples of each class of profile: Type I, Type $\rm I_o$, Type $\rm II_o$,
and Type $\rm III_o$ ({\em top to bottom}). {\em Left panels:} ACS $V$-band images. {\em Right panels:}
Azimuthally averaged $V$-band radial surface brightness profiles. We overplot the break radii
in the outer stellar disc, $24 < \mu < 26.5\,\rm mag\,arcsec^{-2}$ ({\em red solid lines}). Also shown
are break radii with $\mu_{\rm brk} < 24\,\rm mag\,arcsec^{-2}$ ({\em red dashed lines}), which are not
considered in this study, and the stellar disc limit $a_{\rm disc\,lim}$ ({\em black dashed line}).
The inner and outer scalelength, $h_{\rm in}$ and $h_{\rm out}$ respectively, and the break strength $T$ are
also shown for reference. The ACS images are in a logarithmic greyscale.}
\end{figure*}

We classify our galaxies into three main types, those classified to be Type $\rm I_o$, Type $\rm II_o$, or
Type $\rm III_o$ depending on break features in their outer stellar disc (o - outer, $24 < \mu_V < 26.5\,{\rm
mag\,arcsec}^{-2}$) only. The classification used assumes only one break in the outer disc. If two breaks
are present the outer break is used for classification. This is only the case for $4$ ($\sim3$ per
cent) field galaxies. If the profile has no break in the outer disc then the galaxy has a simple
exponential outer disc and is classified as Type $\rm I_o$. Galaxies that are pure
exponentials across the length of their surface brightness profile, aside from the varying bulge component
(Type 1, a subset of Type $\rm I_o$), are also identified. If the profile has a down-break in their outer
$\mu$ profile then the galaxy has a truncation in the outer disc and is classified as Type $\rm II_o$.
However, if the profile has an up-break then the galaxy has an antitruncation in the outer disc and is
classified as Type $\rm III_o$.  Examples of each profile type (Type I, $\rm I_o$, $\rm II_o$, and $\rm
III_o$) are shown in Fig.~\ref{Profile types} along with their ACS images showing the break radius $r_{\rm
brk}$ isophote. Please note that this classification scheme is different to that used by previous works
\cite[e.g.][]{Pohlen_Trujillo:2006}.

\subsection{Determining scalelengths and break strength}

\label{Determining scalelengths and break strength}

In the following, our exponential fits are obtained using a linear least-squares fit to the original
$\mu$ profile between radii identified during the visual inspection (see Section~\ref{Profile inspection}).

For galaxies with no break in their outer disc (Type $\rm I_o/I$, pure exponentials) we obtain the
scalelength $h$ using a simple exponential fit ($h = 1.086\times\Delta{r}/\Delta{\mu_{\rm fit}(r)}$) across
the length of the disc component. For bulgeless spiral galaxies with pure exponential discs (Type I Sd
galaxies), the scalelength $h$ was confirmed to be compatible with the {\sc galfit} effective radius $r_e$
($r_e\sim1.7h$). The mean random error in scalelength due to the exponential fitting routine is
$<10$ per cent for these galaxies. The mean systematic error in the scalelength due to the error in the sky
subtraction $\pm1\sigma$ (see Section \ref{Sky subtraction}) is also $<10$ per cent.

For galaxies where an outer disc profile break was identified (Type $\rm II_o/III_o$), we obtain the
scalelength {\em h} of exponential fits either side of the break radius $r_{\rm brk}$, and therefore obtain
an inner and outer exponential fit for the stellar disc. The inner exponential disc extends from a radius of
$r_{\rm in,min}$ to $r_{\rm in,max}$ and has a scalelength $h_{\rm in}$ given by

\begin{equation}
h_{\rm in} = 1.086\times\frac{r_{\rm in,max}-r_{\rm in,min}}{\mu_{\rm fit}(r_{\rm in,max})-\mu_{\rm fit}
(r_{\rm in,min})},
\end{equation}

where $\mu_{\rm fit}$ is the surface brightness from the exponential fit. Similarly, the outer exponential
disc extends from a radius of $r_{\rm out,min}$ to $r_{\rm out,max}$ and has a scalelength $h_{\rm out}$
given by

\begin{equation}
h_{\rm out} = 1.086\times\frac{r_{\rm out,max}-r_{\rm out,min}}{\mu_{\rm fit}(r_{\rm out,max})-\mu_{\rm fit}
(r_{\rm out,min})}.
\end{equation}

For these galaxies, the mean random error in scalelength due to the exponential fitting routine is
$<10$ per cent for $h_{\rm in}$ and $<20$ per cent for $h_{\rm out}$. In order to measure the strength of
the profile break $T$ we use the logarithm of the outer-to-inner scalelength ratio,

\begin{equation}
T = {\rm log}_{10}\,h_{\rm out}/h_{\rm in}.
\end{equation}

\begin{figure}
\includegraphics[width=0.47\textwidth]{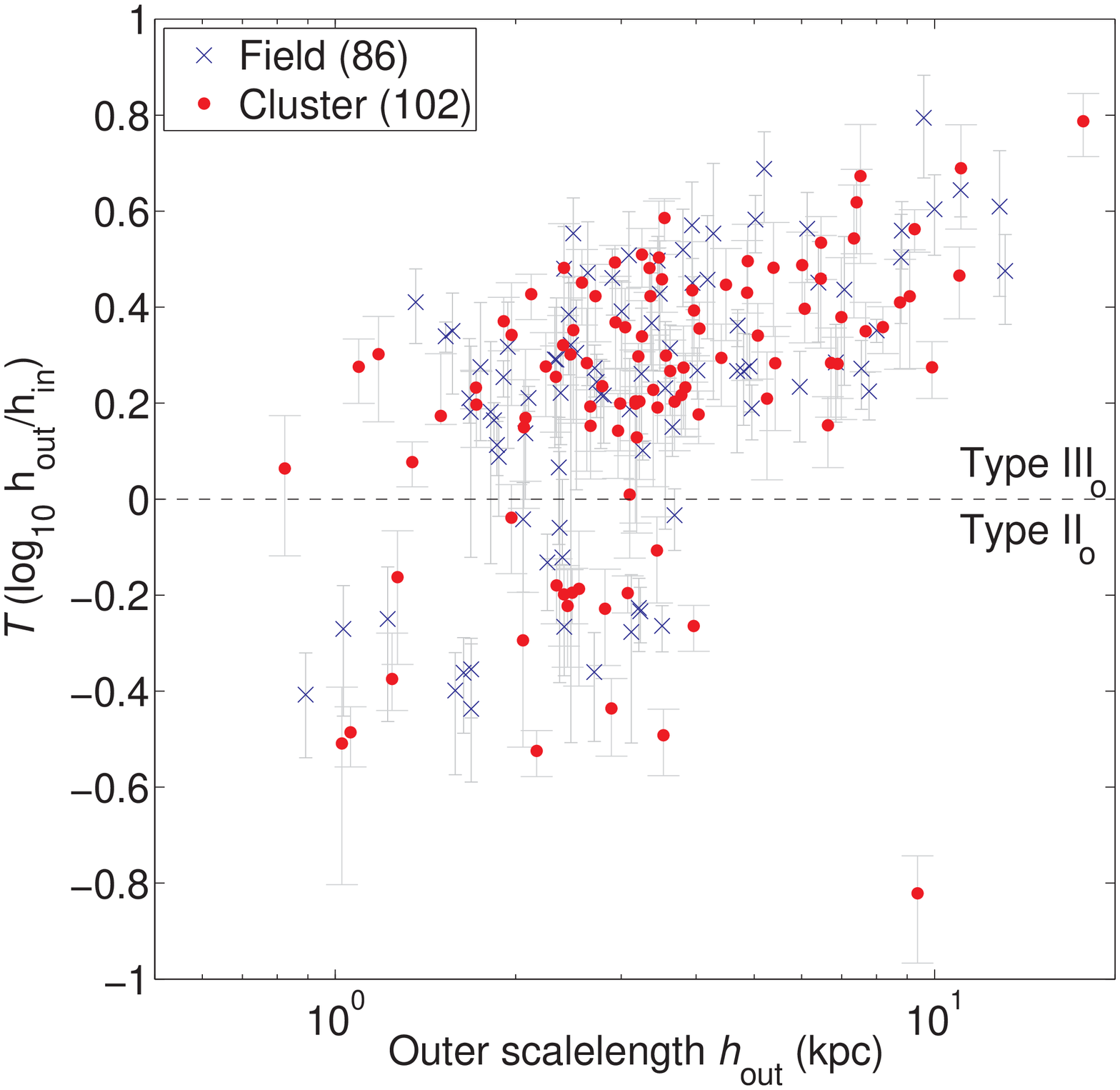}
\caption{\label{T error} A comparison of break strength $T$ and outer scalelength $h_{\rm out}$ for our Type
$\rm II_o/III_o$ galaxies in the field ({\em blue crosses}) and cluster ({\em red points}) environment (as
classified by DTM). An outlying cluster galaxy located at $h_{\rm out} = 453.2\rm\,kpc$, $T = 2.382$ is not 
shown for clarity. The $T$ errorbars represent the error from the sky subtraction ($\pm1\sigma$) on a
galaxy-galaxy basis. The mean error in $T$ is $\pm0.1$. Respective sample sizes are shown in the legend.}
\end{figure}

The mean random error in $T$ due to the exponential fitting routine is $\sim\pm0.1$ ($<20$ per cent).
Fig.~\ref{T error} shows the break strength $T$ plotted against the outer scalelength $h_{\rm out}$ for Type
$\rm II_o/III_o$ galaxies (classified by DTM) with the errorbars in $T$ due to the sky subtraction error
$\pm1\sigma$ (see Section \ref{Sky subtraction}). The mean systematic error in $T$ due to the sky
subtraction error is also $\sim\pm0.1$.

\begin{figure*}
\includegraphics[width=0.99\textwidth]{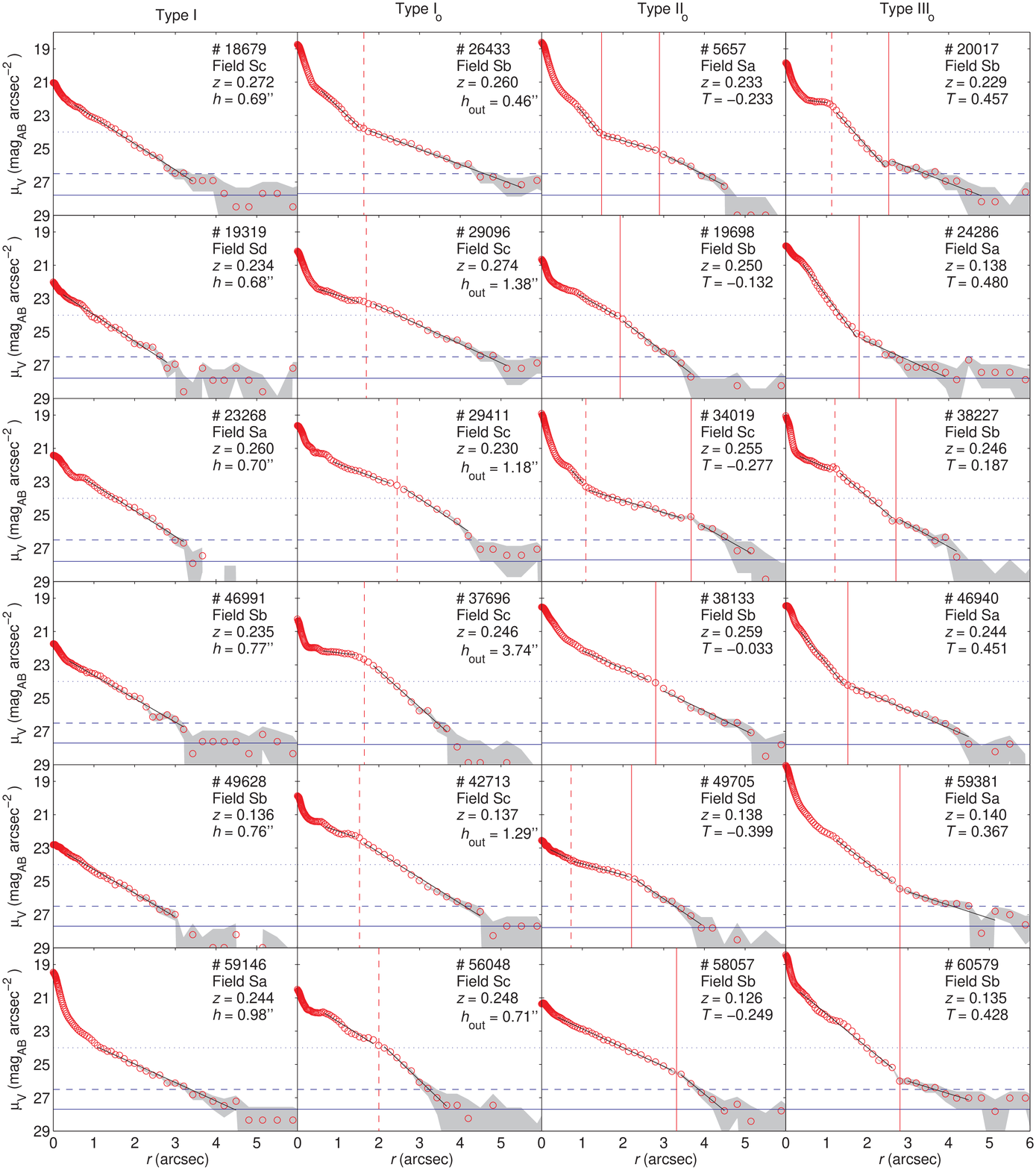}
\caption{\label{Example field profiles} Example azimuthally averaged radial surface brightness profiles for
different profile types in the field environment. {\em First column:} Type I profiles (pure exponential
profiles). {\em Second column:} Type $\rm I_o$ profiles which have a simple exponential region in the outer
disc ($24 < \mu_V < 26.5\,\rm mag\,arcsec^{-2}$, region between blue dashed and dotted lines). {\em
Third column:} Type $\rm II_o$ profiles (down-bending break in outer disc, {\em outer disc truncations}).
{\em Forth column:} Type $\rm III_o$ profiles (up-bending break in outer disc, {\em outer disc
antitruncations}). The error in the $\mu$ profiles is for an over- and undersubtraction of the sky by
$\pm1\sigma$. The critical surface brightness $\mu_{\rm crit}$ ($\pm1\sigma$ sky error, {\em blue solid line})
and limiting surface brightness $\mu_{\rm lim}$ ($\pm3\sigma$ sky error, {\em blue dashed line}) are shown on
the profiles. The position of outer disc profile breaks ($24 < \mu_{\rm brk} < 26.5\,\rm mag\,arcsec^{-2}$)
are overplotted as {\em red solid lines}, breaks with $\mu_{\rm brk} < 24\,\rm mag\,arcsec^{2}$ are represented
by {\em red dashed lines}. The least square exponential fits to exponential regions are overplotted on the
relevant sections of the surface brightness profiles. The Hubble type, redshift $z$, scalelength $h$, and break
strength $T$ of the galaxies are also shown for reference. The ACS images and $\mu$ profiles for all galaxies
in our samples are presented in Appendix~A (online version only).}
\end{figure*}

\begin{figure*}
\includegraphics[width=0.99\textwidth]{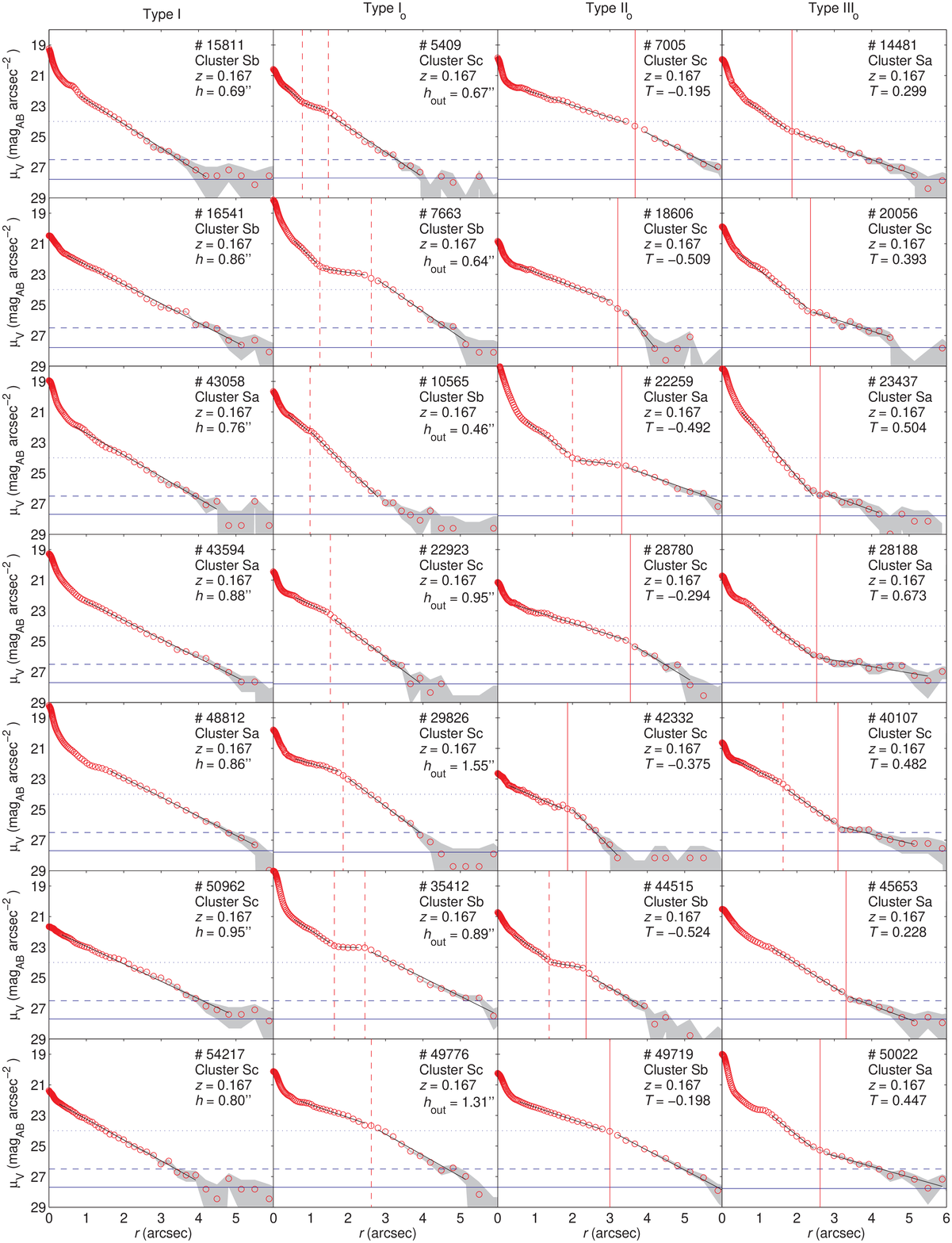}
\caption{\label{Example cluster profiles} Example azimuthally averaged radial surface brightness profiles
for different profile types in the cluster environment. Figure same as Fig.~\ref{Example field profiles} but
for the cluster environment. The ACS images and $\mu$ profiles for all galaxies in our samples are presented
in Appendix~A (online version only).}
\end{figure*}

A Type $\rm I_o$/I galaxy (pure exponential) has no break, and therefore has a break strength of $T = 0$. A
Type $\rm II_o$ galaxy (down-bending break, truncation) has a smaller outer scalelength $h_{\rm out}$ with
respect to its inner scalelength $h_{\rm in}$, and therefore has a negative break strength ($T < 0$).
Similarly, a Type $\rm III_o$ galaxy (up-bending break, antitruncation) has a larger outer scalelength
$h_{\rm out}$ with respect to its inner scalelength $h_{\rm in}$, and therefore has a positive break
strength ($T > 0$). We present a selection of our $\mu$ profiles with fitted exponential regions and
overplotted break radii from both the field and cluster environments in Figs.~\ref{Example field profiles}
and \ref{Example cluster profiles} respectively. The ACS images and $\mu$ profiles for all galaxies in our
samples are presented in Appendix~A (online version only).

In the outer regions of the surface brightness profile, negative sky-subtracted flux can occur.
Surface brightness cannot be defined for a negative flux. Therefore, these points are removed from our linear
exponential fits to surface brightness. However, this could potentially lead to a slight measurement bias and
a larger outer scalelength $h_{\rm out}$ for some profiles. In order to address this issue we repeated all
measurements of scalelength ($h_{\rm in}$ and $h_{\rm out}$) and break strength $T$ using a non-linear least
squares exponential fit to the flux radial profile (retaining negative fluxes). We then compared these
results to those from the linear fits to surface brightness in order to assess if any measurement bias was
present. The results from both fitting methods were in very good agreement and no bias was observed.
Therefore, we use the linear fits to surface brightness in our analysis.

\section[]{Results}

\label{Results}

The frequency of profile classifications by each assessor (DTM, AAS, MEG) for spiral galaxies in the field
and cluster environments are shown in Table~\ref{Classifications tbl}. In each case the profile
classification is based on $\mu_V$ breaks in the outer stellar disc ($24 < \mu_V < 26.5\,{\rm mag\,
arcsec}^{-2}$). Uncertainty in the frequency/fraction of profile types $\delta{f}$ ($f = N_{i}/N_{\rm
total}$) are calculated using

\begin{equation}
\left ( \frac {\delta{f_{i}}}{f_{i}} \right )^2 = \left ( \frac {\delta{N_{i}}}{N_{i}} \right )^2 + \left
( \frac {\delta{N_{\rm total}}}{N_{\rm total}} \right )^2 - \frac {2 \sqrt{N_{i}N_{\rm total}}}{N_{i}N_{\rm
total}},
\label{error_eqn}
\end{equation}

where $\delta{N_{i}} = \sqrt{N_{i}}$ and $\delta{N_{\rm total}} = \sqrt{N_{\rm total}}$.

\begin{table*}
\begin{minipage}{175mm}
\centering
\caption{\label{Classifications tbl}{The frequency of profile types in the field and cluster environments
for the three independent assessors (DTM, AAS, MEG). Type I is a sub-sample of Type $\rm I_o$ and
percentage errors are calculated using equation \ref{error_eqn}.}}
\begin{tabular}{lcccccc}
\hline
\hline
{Assessor}		&{Type I}		&\multicolumn{3}{c}{Outer disc profile types}	&{Unclassified}	\\	
{}			&{}			&{Type $\rm I_o$}	&{Type $\rm II_o$}	&{Type $\rm III_o$}	&{}	\\
\hline
{Field galaxies}	&{}			&{}			&{}			&{}			&{}			\\
{DTM}			&{$17$ ($12\pm2\,\%$)}	&{$59$ ($41\pm2\,\%$)}	&{$18$ ($12\pm2\,\%$)}	&{$68$ ($47\pm2\,\%$)}	&{$0$}			\\
{AAS}			&{$42$ ($29\pm3\,\%$)}	&{$76$ ($52\pm2\,\%$)}	&{$14$ ($10\pm2\,\%$)}	&{$55$ ($38\pm2\,\%$)}	&{$0$}			\\
{MEG}			&{$38$ ($26\pm3\,\%$)}	&{$86$ ($59\pm2\,\%$)}	&{$16$ ($11\pm2\,\%$)}	&{$40$ ($28\pm3\,\%$)}	&{$3$ ($2\pm2\,\%$)}
\vspace{2mm} 																	\\
{Cluster galaxies}	&{}			&{}			&{}			&{}			&{}			\\
{DTM}			&{$28$ ($15\pm2\,\%$)}	&{$80$ ($44\pm2\,\%$)}	&{$19$ ($10\pm2\,\%$)}	&{$83$ ($46\pm2\,\%$)}	&{$0$}			\\
{AAS}			&{$48$ ($26\pm2\,\%$)}	&{$83$ ($46\pm2\,\%$)}	&{$23$ ($13\pm2\,\%$)}	&{$74$ ($41\pm2\,\%$)}	&{$2$ ($1\pm1\,\%$)}	\\
{MEG}			&{$48$ ($26\pm2\,\%$)}	&{$88$ ($48\pm2\,\%$)}	&{$17$ ($9 \pm2\,\%$)}	&{$69$ ($38\pm2\,\%$)}	&{$8$ ($4\pm2\,\%$)}	\\
\hline
\hline
\end{tabular}
\end{minipage}
\end{table*}

Due to the subjective nature of some profile classifications, the frequency obtained for each profile type
varies subtly between the different assessors. The agreement between the three assessors is generally good
for the outer disc profile types (Type $\rm I_o$, $\rm II_o$, and $\rm III_o$). This agreement is very good
in the cluster, but slightly weaker in the field due to most of the galaxies being at higher redshift and
therefore having poorer quality profiles. There is less agreement between the three assessors for the
frequency of Type I profiles (pure exponentials) due to the subjective nature of the classification of weak
profile breaks across the length of the $\mu$ profile and inner profile breaks near the bulge component.

The frequency of profile types in the outer stellar disc (Type $\rm I_o$, $\rm II_o$, and $\rm III_o$) are
approximately the same in the field and cluster environment. For both field and cluster spirals, $\sim50$ per
cent have a simple exponential profile in the outer stellar disc (Type $\rm I_o$), $\sim10$ per cent exhibit
a down-bending break (outer disc truncation, Type $\rm II_o$), and $\sim40$ per cent exhibit an up-bending
break (outer disc antitruncation, Type $\rm III_o$). The frequency of Type I profiles is also the same in the
field and cluster for each assessor and is $\sim20\pm10$ per cent. We stress that
due to our profile classification being based on breaks with $\mu_{\rm brk} > 24 \,{\rm mag\,arcsec}^{-2}$,
our profile type fractions do not necessarily need to agree with those of previous works \cite[e.g.][]{Pohlen_Trujillo:2006}.
For our cluster galaxies, we also investigated
the frequency of outer disc profile types as a function of
stellar surface mass density (range $10^{10}$--$10^{13}\rm\,M_\odot\,Mpc^{-2}$, see \cite{Wolf_etal:2009} for
details of how this was measured). However, no significant differences were observed. These results suggest
that the profile type in the outer disc of spiral galaxies is not affected by the environment from the field
out to the intermediate densities of the A901/2 clusters.

In this work, we define the outer disc to be $24 < \mu < 26.5\,\rm mag\,arcsec^{-2}$, see
Section~\ref{Defining the Outer Stellar Disc}. Our outer disc profile types are based on this somewhat
arbitrary but justified definition. Independently and uniformly adjusting the upper and lower $\mu$ limits for
the outer disc by $+0.5\,\rm mag\,arcsec^{-2}$ affects our profile type fractions by $<10$ per cent. In each case,
there remains no significant differences between the fractions in the field and cluster environment.

The majority of our field sample have redshift $z\sim0.23$ while the cluster galaxies have $z\sim0.167$.
However, evolutionary effects are not expected to have an impact on the results of this work. The break
strength $T$ of our field galaxies show no correlation with redshift and evolutionary effects on the scalelength
$h$ between the mean redshifts of our field and cluster samples is only $\sim5$ per cent (based on the fits of
\citealt{Buitrago_etal:2008} for the expected size evolution of massive disc galaxies).

\subsection{Morphology}

\label{Morphology}

\begin{figure}
\includegraphics[width=0.33\textwidth]{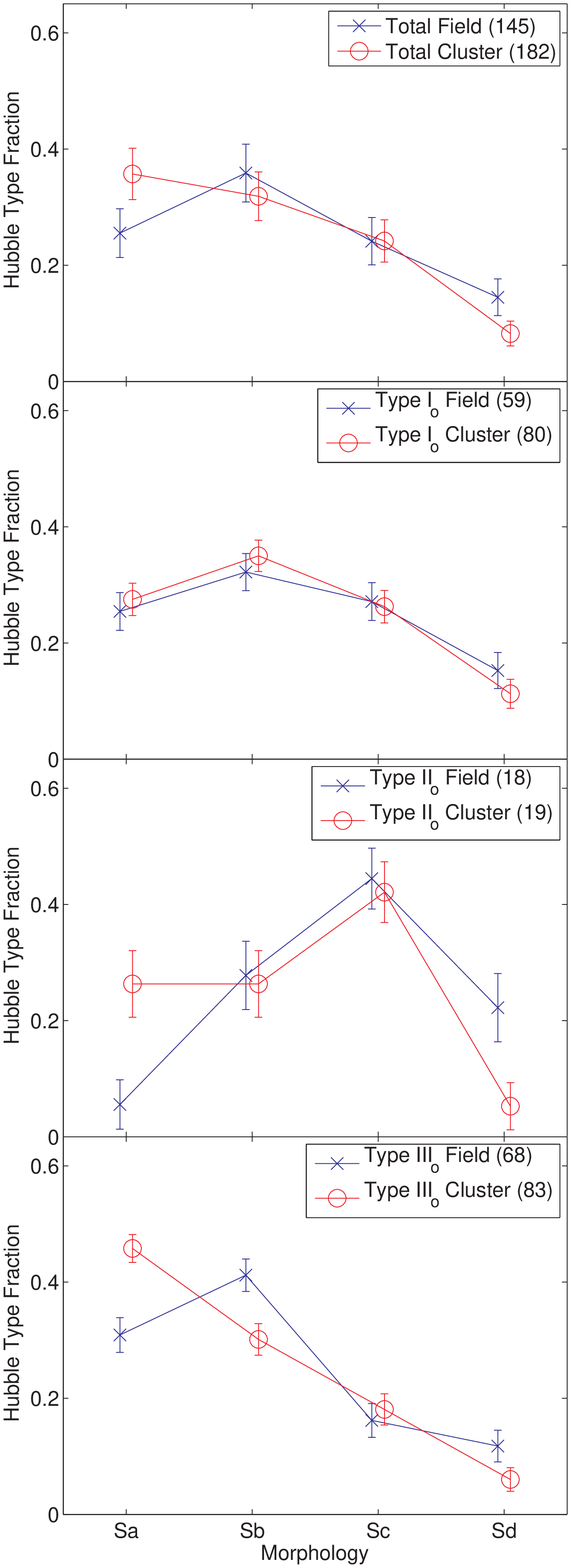}
\centering
\caption{\label{Morphology mix per type} The morphological mix for outer disc profile types. The fraction of
each Hubble type in the field ({\em crosses}) and cluster ({\em circles}) environment for ({\em top to bottom})
the total spiral galaxy sample, Type $\rm I_o$ galaxies, Type $\rm II_o$ galaxies, and Type $\rm III_o$
galaxies. Respective sample sizes are shown in the legends and the errors in the Hubble Type fraction were
calculated using equation \ref{error_eqn}. Contamination of the cluster sample by the field is $< 25$ per cent.
We observe no significant difference between the morphological mix in the field and cluster environment for
each profile type.}
\end{figure}

\begin{figure}
\includegraphics[width=0.33\textwidth]{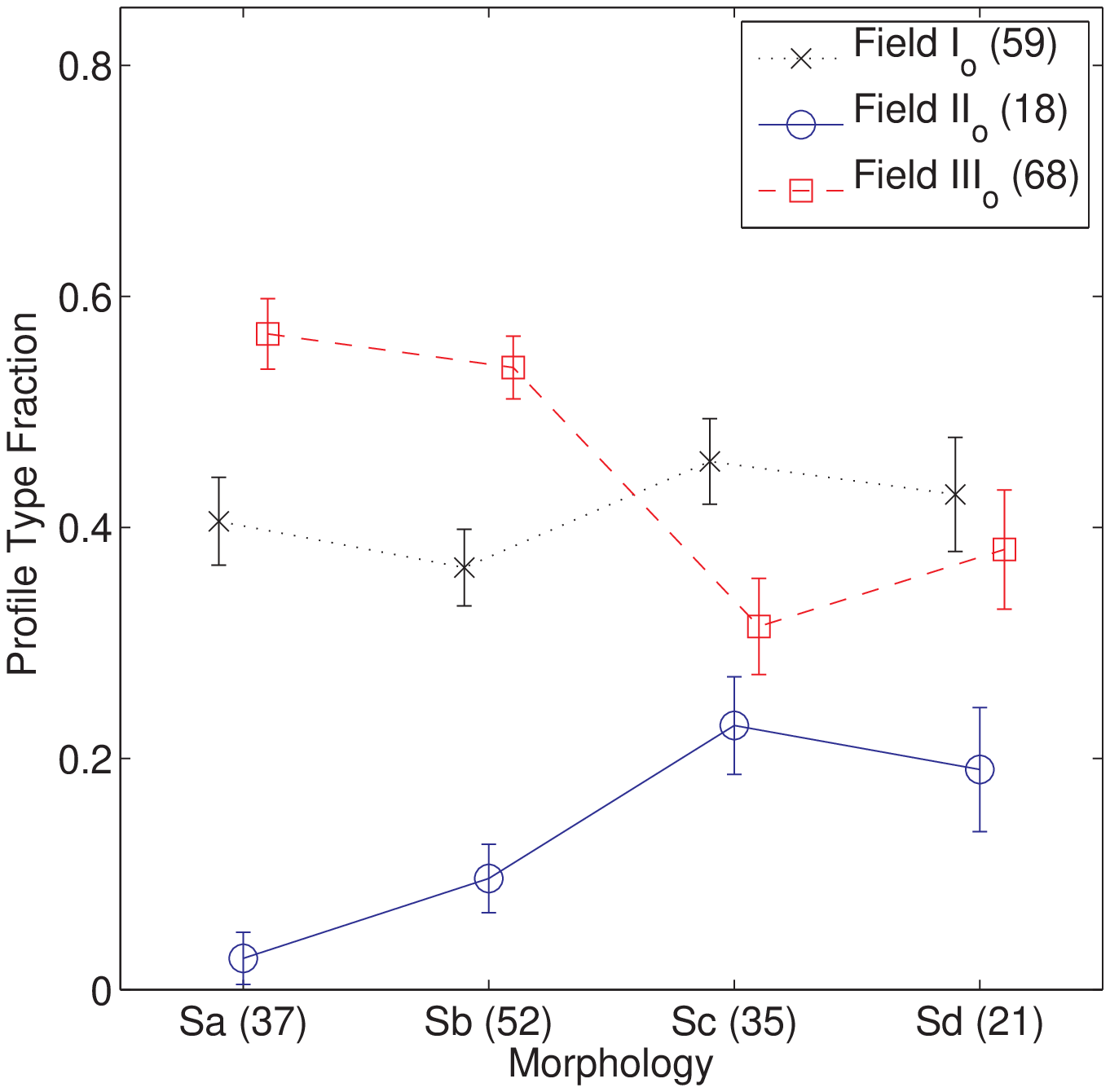} \\
\includegraphics[width=0.33\textwidth]{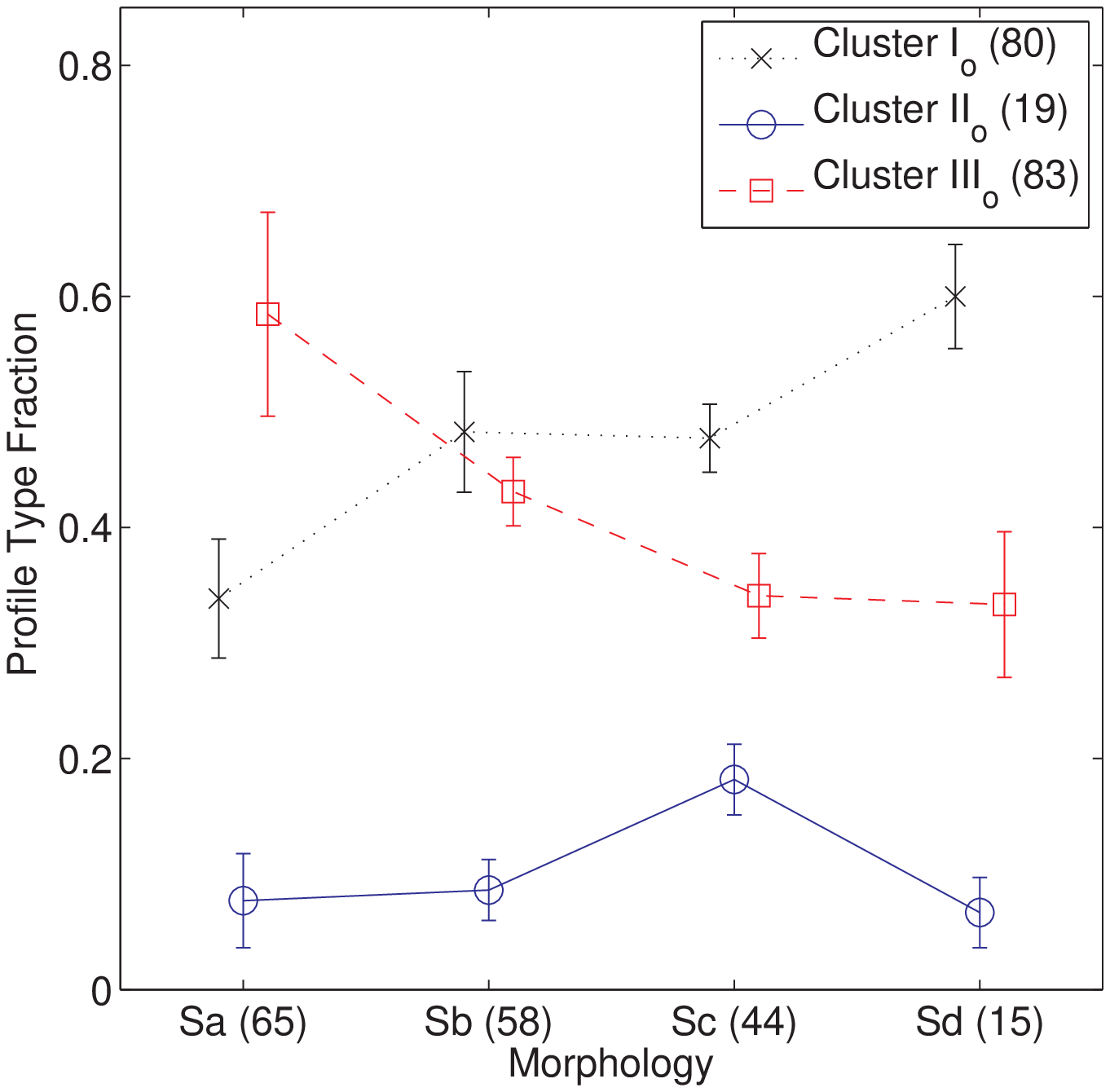} \\
\centering
\caption{\label{Type mix per morphology} The fraction of outer disc profile types, Type $\rm I_o$, ({\em
crosses}), Type $\rm II_o$ ({\em circles}), and Type $\rm  III_o$ ({\em squares}) for different spiral
Hubble type morphologies in the field ({\em top}) and cluster ({\em bottom}) environment. The
sample size for each profile type is shown in the legends and the sample size for each Hubble type is shown
on the $x$ axis. The errors in the profile type fraction were calculated using equation \ref{error_eqn}.
Contamination of the cluster sample by the field is $< 25$ per cent. We observe similar trends for both field
and cluster galaxies. The fraction of Type $\rm I_o$ is approximately constant with Hubble type, perhaps rising
slightly towards later Hubble types in the cluster. However, the fraction of Type $\rm II_o$ galaxies
(truncations) increases slightly towards later Hubble types while the fraction of Type $\rm III_o$ galaxies
(antitruncations) decreases.}
\end{figure}

For each outer disc profile type (Type $\rm I_o$, $\rm II_o$, and $\rm III_o$), we compare the morphological
mix (Sa, Sb, Sc, Sd) in the field and cluster environment to that of the total spiral galaxy sample (see
Fig.~\ref{Morphology mix per type}). We also considered whether or not the galaxies were barred.

The morphological mix of the total galaxy sample is approximately the same in the field and cluster, with
perhaps a slight excess of Sa galaxies in the cluster environment (Sa galaxies are visually defined to have
a high bulge-to-disc ($B/D$) ratio, see Section~\ref{Description of the data}). This excess is dominated by
barred Sa galaxies. For each outer disc profile type (Type $\rm I_o$, $\rm II_o$, and $\rm III_o$), the
morphological mix is also approximately the same in the field and cluster environment. However, there is a
potential slight excess of Sa galaxies in the cluster for both Type $\rm II_o$ and $\rm III_o$ galaxies.
In both cases the excess is dominated by barred Sa galaxies. Except for the slight excess of barred Sa
galaxies in the cluster, the morphological mix for barred and unbarred galaxies are the same in the field
and cluster for each profile type.

The morphological mix of Type $\rm I_o$ galaxies is comparable to that of the total galaxy sample. This
implies that Type $\rm I_o$ galaxies are equally probable for all morphological types. For Type $\rm II_o$
galaxies, the morphological mix fractions are affected by low number statistics. However, there is a clear
trend with the Hubble type fraction (Sa--Sc) rising towards later Hubble types. For Type $\rm III_o$
galaxies, the morphological mix is also comparable to that of the total galaxy sample. There is also a clear
trend with the Hubble type fraction (Sa--Sd) decreasing towards later Hubble types.

We conclude that outer disc truncations (Type $\rm II_o$) are more prevalent in later Hubble types while
outer disc antitruncations (Type $\rm III_o$) are more prevalent in early Hubble types. However, there is
no evidence to suggest the morphological mix for the different outer disc profile types is affected by the
galaxy environment.

We also compare the frequency of outer disc profile types (Type $\rm I_o$, $\rm II_o$, and $\rm III_o$) for
different Hubble type morphologies (Sa, Sb, Sc, Sd) in both the field and cluster environment (see
Fig.~\ref{Type mix per morphology}). For Type $\rm I_o$ galaxies, we observe no clear correlation between the
frequency of profile type and the Hubble type in both the field and cluster environment. The frequency of Type
$\rm I_o$ galaxies is approximately the same for each Hubble type morphology, perhaps raising slightly towards
later Hubble types in the cluster environment. For Type $\rm II_o$ and Type $\rm III_o$ galaxies, a weak trend
is observed between the frequency of profile type and the Hubble type in both the field and cluster
environments. The frequency of Type $\rm II_o$ galaxies increases slightly towards later Hubble types while
the frequency of Type $\rm III_o$ galaxies decreases. For Sd galaxies, low number statistics could be masking
the continuation of these trends. These results are in qualitative agreement with the work of
\cite{Pohlen_Trujillo:2006} who find a clear correlation between the frequency of profile type and Hubble type
using a sample of $\sim90$ nearby galaxies from SDSS (classification based on $\mu$ breaks across the entire
stellar disc so direct comparisons cannot be made). We conclude that there is no significant difference
between the frequency of outer disc profile types for different Hubble type morphologies in the field
and cluster environments.

\subsection[]{Pure exponential outer discs (Type $\rm I_o$)}

\label{Type I/Io analysis}

\begin{figure*}
\includegraphics[width=1\textwidth]{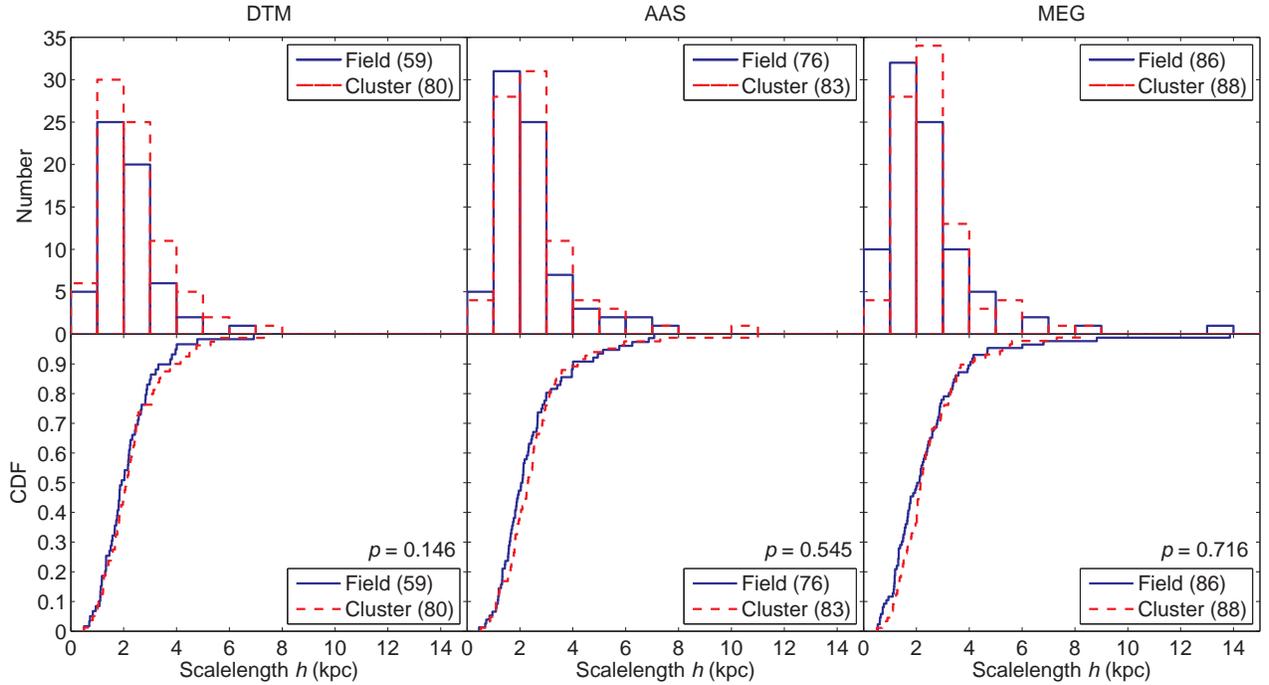}
\caption{\label{Type Io analysis} Comparing outer disc scalelength $h_{\rm out}$ distributions in different
environments. {\em Top row:} Outer disc scalelength $h_{\rm out}$ distributions for Type $\rm I_o$ galaxies
in the field ({\em blue line}) and cluster ({\em red dashed line}) environment as classified by DTM ({\em
left panel}), AAS ({\em centre panel}), and MEG ({\em right panel}). {\em Bottom row:} The corresponding
outer scalelength $h_{\rm out}$ CDFs showing the probability $p$ that compared samples are {\em not} drawn from
the same continuous $h_{\rm out}$ distributions in the bottom right of each plot. Respective sample sizes
are shown in the legends. Random errors in scalelength are typically $<10$ per cent. Systematic errors in
scalelength due to the error in the sky subtraction are also typically $<10$ per cent. Contamination of the
cluster sample by the field is $< 25$ per cent. We find no significant difference between the CDFs in each
environment and no evidence to suggest that the outer scalelength $h_{\rm out}$ of our Type $\rm I_o$ galaxies
are {\em not} drawn from the same continuous $h_{\rm out}$ distributions.}
\end{figure*}

For galaxies where no outer disc $\mu$ break was identified (Type $\rm I_o$ galaxies, which include all Type
I galaxies), we compare the outer scalelength $h_{\rm out}$ distributions in the field and cluster
environments to see if there is any evidence for an environmental dependence on outer scalelength $h_{\rm
out}$ (see Fig.~\ref{Type Io analysis}). The outer scalelengths $h_{\rm out}$ for these Type $\rm I_o$
galaxies were transformed into intrinsic linear scales using a cosmology of $H_0 = 70\, {\rm kms^{-1}
Mpc^{-1}}$, $\Omega_\Lambda = 0.7$, and $\Omega_m = 0.3$. We use the fixed cluster redshift ($z = 0.167$) to
determine the intrinsic outer scalelengths of our cluster Type $\rm I_o$ galaxies and the original COMBO-17
redshift estimate for our field Type $\rm I_o$ galaxies. Therefore, photo-$z$ errors only propagate into the
intrinsic scalelengths of our field galaxies and not our cluster galaxies. The mean error in $h_{\rm out}$
associated with the photo-$z$ error is $<10$ per cent \citep{Maltby_etal:2010}.

In our surface brightness $\mu$ profiles, the error in the {\sc galapagos} sky (see Section~\ref{Sky
subtraction}) can have a significant effect on the scalelength $h$ and break strength $T$, especially in the
outer regions where the surface brightness $\mu$ approaches the critical surface brightness $\mu_{\rm crit}$
($27.7\,{\rm mag\, arcsec}^{-2}$). However, for any particular galaxy the sky subtraction error can be taken
to be approximately constant across the length of the surface brightness profile. Therefore, we can account
for this error by performing parallel analysis for when the sky background is over- and undersubtracted by
$\pm1\sigma$. The mean error in $h_{\rm out}$ due to the sky subtraction error is $\sim\pm 0.3\,\rm kpc$
($<10$ per cent) for our Type $\rm I_o$ galaxies. Random errors in $h_{\rm out}$ due to exponential
fitting are also typically $<10$ per cent (see Section \ref{Determining scalelengths and break strength}). We
also perform parallel analysis on the Type $\rm I_o$ samples generated by the three assessors (DTM, AAS, MEG)
in order to account for the subjective nature of the profile classifications and compare the final results.

In all cases (all parallel analyses), we observe no clear difference between the distributions of outer
disc scalelength $h_{\rm out}$ for our Type $\rm I_o$ spiral galaxies in the field and cluster
environments (see Fig.~\ref{Type Io analysis}).

In order to test the significance of these results we construct outer disc scalelength $h_{\rm out}$
cumulative distribution functions (CDFs, see Fig~\ref{Type Io analysis}) for the Type $\rm I_o$ galaxy
samples and perform Kolmogorov--Smirnov (K--S) tests between corresponding samples from the field and
cluster environments. These K--S tests are used in order to obtain the probability that the field and
cluster Type $\rm I_o$ samples are {\em not} drawn from the same continuous $h_{\rm out}$ distributions. The results
of these K--S tests are shown in Table~\ref{K-S results tbl}.

In this work, we only consider an environmental effect on the Type $\rm I_o$ outer scalelength $h_{\rm out}$
to be significant if K--S tests yield a $2\sigma$ level probability for the field and cluster Type $\rm I_o$
galaxy samples {\em not} being drawn from the same continuous $h_{\rm out}$ distributions. This probability $p{(\rm
Field/Cluster)}$ is below the $2\sigma$ level for each assessor and for when the sky background is over- and
undersubtracted by $\pm1\sigma$. Therefore, we find no evidence to suggest that the outer disc scalelength
$h_{\rm out}$ of Type $\rm I_o$ galaxies is dependent on the galaxy environment.

We perform the same analysis for Type I galaxies, a sub-sample of Type $\rm I_o$ galaxies. These galaxies
have simple exponential $\mu$ profiles across the length of their disc component. The K--S test results
for these galaxies are shown in Table~\ref{K-S results tbl}. The probability $p{(\rm Field/Cluster)}$ is
below the $2\sigma$ level in each case. Therefore, we also find no evidence to suggest that the scalelength
$h$ of Type I galaxies is dependent on the galaxy environment.

\subsection[]{Outer disc breaks (Type $\rm II_o/III_o$)}

\label{Type IIo/IIIo analysis}

\begin{figure*}
\includegraphics[width=1\textwidth]{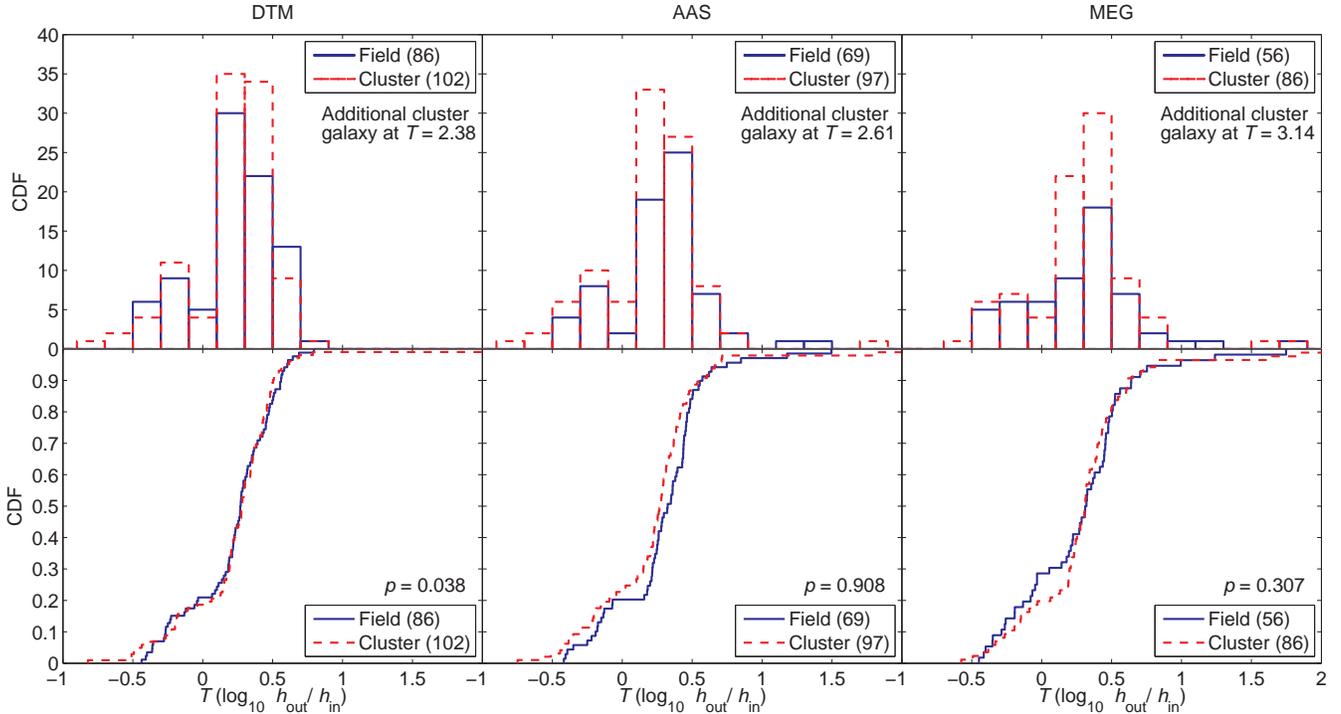}
\centering
\caption{\label{Type II/III analysis} Comparing break strength $T$ (${\rm log}_{10}\,h_{\rm out}/{h}_{\rm
in}$) distributions in different environments. {\em Top row:} Break strength $T$  distributions for Type
$\rm II_o/III_o$ galaxies in the field ({\em blue line}) and cluster ({\em red dashed line}) environment as
classified by DTM ({\em left column}), AAS ({\em centre column}), and MEG ({\em right column}). {\em Bottom
row:} The corresponding break strength $T$ CDFs showing the probability $p$ that compared samples are {\em not}
drawn from the same continuous $T$ distributions in the bottom right of each plot. Respective sample sizes are
shown in the legends. Random errors in break strength $T$ are typically $<0.1$. Systematic errors in break
strength due to the error in the sky subtraction are also $\sim\pm0.1$. Contamination of the cluster sample by
the field is $< 25$ per cent. We find no significant difference between the CDFs in each environment and no
evidence to suggest that the break strength $T$ of our Type $\rm II_o/III_o$ galaxies are {\em not} drawn from
the same continuous $h_{\rm out}$ distributions.}
\end{figure*}

For our Type $\rm II_o$ and $\rm III_o$ galaxies, we compare the distribution of break strength $T$ in the
field and cluster environments to see if there is any evidence for an environmental dependence on break
strength $T$ in the outer disc (see Fig.~\ref{Type II/III analysis}). We perform similar parallel analyses
and statistical tests as for our Type $\rm I_o$ galaxies. The mean error in $T$ due to the sky subtraction
error is $\pm0.1$. Random errors in $T$ due to exponential fitting are also typically $\pm0.1$ (see Section
\ref{Determining scalelengths and break strength}). The results of the K--S tests are shown in
Table~\ref{K-S results tbl}. 

In all cases (all assessor samples and sky versions), we observe no clear difference between the $T$
distributions in the field and cluster environments.  The probability $p({\rm Field/Cluster})$ is below the
$2\sigma$ level in each case. Therefore, we find no evidence to suggest that the break strength $T$ of our
Type $\rm II_o/III_o$ galaxies is dependent on the galaxy environment and this result is robust to the error
in the sky background and the subjective nature of the profile classifications.

We explore this result by splitting the combined field and cluster Type $\rm II_o/III_o$ galaxy sample into
two sub-samples about the median value for stellar mass $M_{*{\rm ,mdn}}$ ($10^{10}\,\rm M_\odot$). Similar
statistical tests are then used to determine the significance of an environmental dependence independently
for high- ($M_* > M_{*{\rm,mdn}}$) and low-mass ($M_* < M_{*{\rm ,mdn}}$) Type $\rm II_o/III_o$ galaxies.
A similar test was also performed using $B-V$ colour instead of stellar mass where $B$ and $V$ were obtained
from the STAGES master catalogue \citep{Gray_etal:2009} and $(B-V)_{\rm,mdn} = 0.70\,{\rm mag}$. In all cases
(all assessor samples and sky versions), the significance of an environmental dependence on break strength $T$
remains below the $2\sigma$ level. Therefore, we conclude that there is no evidence to suggest an environmental
dependence on break strength $T$ is dependent on either the stellar mass or the $B-V$ colour of Type
$\rm II_o/III_o$ galaxies.

Another test performed was a comparison of the break radius $r_{\rm brk}$ (in units of the {\sc galfit}
effective radius) in the field and cluster environments for both our Type $\rm II_o$ ($T < 0$) and
Type $\rm III_o$ ($T > 0$) galaxies separately. In both cases, no significant difference ($<2\sigma$) was
observed between the break radius in the field and cluster environments. However, there is a slight
indication that Type $\rm II_o$ breaks occur at smaller break radii than Type $\rm III_o$ breaks.

\begin{figure}
\includegraphics[width=0.45\textwidth]{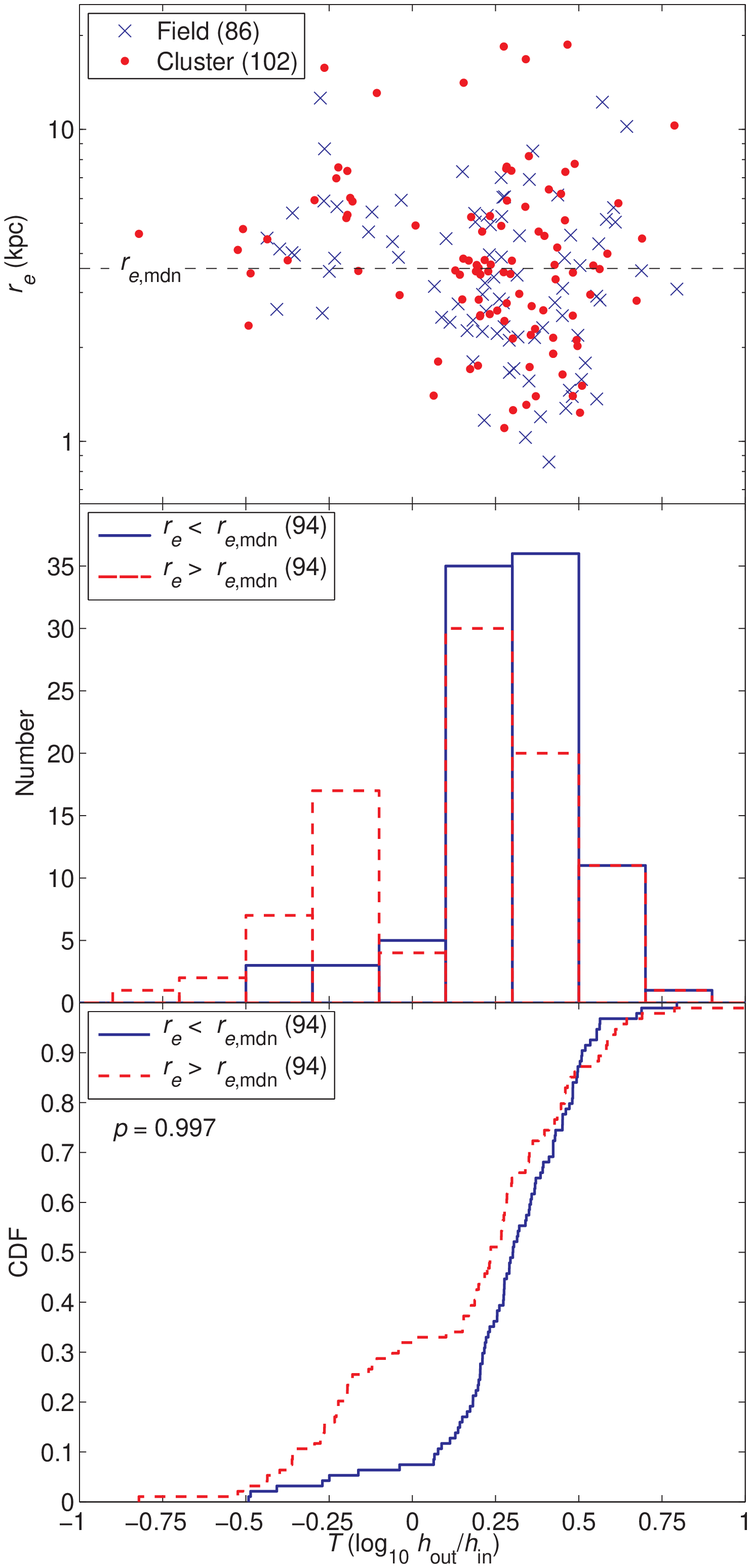}
\centering
\caption{\label{T vs size} A comparison between the break strength $T$ (${\rm log}_{10}\,h_{\rm out}/{h}_{\rm
in}$) and physical size $r_e$ of our Type $\rm II_o/III_o$ galaxies (defined by DTM). {\em Top Panel:} The
effective radius $r_e$--break-strength $T$ distributions for our field ({\em blue crosses}) and cluster
({\em red points}) Type $\rm II_o/III_o$ galaxies, showing the position of the median effective radius
$r_{e{\rm,mdn}}$ ($3.58\,\rm kpc$) of the combined field and cluster sample ({\em black dashed line}).
An outlying cluster galaxy, located at $T = 2.382$, $r_e = 4.442\,\rm kpc$ is not shown for clarity. {\em
Middle panel:} A comparison of the distribution of break strength $T$ for small-$r_e$, $r_e < r_{e{\rm
,mdn}}$ ({\em blue line}) and large-$r_e$, $r_e > r_{e{\rm ,mdn}}$ ({\em red dashed line}) Type $\rm
II_o/III_o$ galaxies. {\em Bottom panel:} A comparison of the break strength $T$ CDFs for small-$r_e$ ({\em
blue line}) and large-$r_e$ ({\em red dashed line}) Type $\rm II_o/III_o$ galaxies. The probability $p$
that small-$r_e$ and large-$r_e$ Type $\rm II_o/III_o$ samples are {\em not} drawn from the same continuous
break strength $T$ distributions is shown in the top left of the plot. Respective sample sizes are shown in the
legend. We find a significant difference between the CDFs of the small-$r_e$ and large-$r_e$ Type $\rm
II_o/III_o$ samples with a high probability (at the $3\sigma$ level) that the samples are {\em not} drawn from
the same continuous break strength $T$ distributions.}
\end{figure}

We also compare the break strength $T$ of our Type ${\rm II_o/III_o}$ galaxies with the effective radius
$r_e$ determined by the STAGES {\sc galfit} S\'{e}rsic models \citep{Gray_etal:2009} (see Fig.~\ref{T vs
size}). The size--break-strength distribution is the same for both field and cluster galaxies. However,
we note that at small effective radii ($r_e < r_{e{\rm ,mdn}}$) there is an absence of Type $\rm II_o$
galaxies (outer disc truncations, $T < 0$) in both the field and cluster environments. $r_{e{\rm ,mdn}}$
($3.58\,\rm kpc$) is the median effective radius of the combined field and cluster Type $\rm II_o/III_o$
galaxy sample. In contrast, there is an abundance of Type ${\rm III_o}$ galaxies (outer disc antitruncations,
$T > 0$) reaching down to $r_e \sim 1\,\rm kpc$. A comparison of the break strength $T$ distributions for
large- ($r_e > r_{e{\rm ,mdn}}$) and small-$r_e$ ($r_e < r_{e{\rm ,mdn}}$) galaxies shows a clear difference
in the two distributions (see Fig.~\ref{T vs size}).

In order to test the significance of this result we construct break strength $T$ CDFs (see Fig.~\ref{T vs
size}), for large- and small-$r_e$ Type $\rm II_o/III_o$ galaxies and perform K--S tests between the large-
and small-$r_e$ galaxy samples to obtain the probability the different samples are {\em not} drawn from the same
continuous break strength $T$ distributions. Similar parallel analyses are also performed as in previous
tests.

In all cases (all assessor samples and sky versions), the probability that the large- and small-$r_e$ galaxy
samples $p$(large-$r_e/$small-$r_e$) are being drawn from different continuous break strength $T$ distributions
is at the $3\sigma$ level. Therefore, we find some evidence to suggest that the break strength $T$ of
Type $\rm II_o/III_o$ galaxies is dependent on the galaxy effective radius $r_e$. We offer no interpretation of
this result.

\subsection[]{Inner/initial disc breaks}
\label{Inner/initial disc breaks}

The focus of this paper is on the effect of the environment on the structure of the outer regions of
galactic discs. By examining only breaks in the outer disc ($\mu > 24 \,{\rm mag\,arcsec}^{-2}$), we have
tailored our analysis to allow for the assessment of an environmental effect on the outer regions of the
galactic disc, where the environment is most likely to have an effect. However, we acknowledge that there is
some potential for missed environmental effects in breaks at a higher surface brightness
($\mu < 24 \,{\rm mag\,arcsec}^{-2}$). Therefore, in order to test whether we could have missed any
environmental effect in the inner disc we repeated our analysis using the inner/initial break for profile
classification. However, we find no correlations of the frequency of profile type with the galaxy environment.
We also find no significant effect of the environment on either the scalelength $h$ of Type I galaxies
(pure exponentials), or the break strength $T$ of inner/initial breaks.


\begin{table}
\begin{minipage}{85mm}
\centering
\caption{\label{K-S results tbl} The K--S test results for Type I, Type $\rm I_o$, and Type $\rm II_o/III_o$
galaxies as classified by DTM, AAS, and MEG. K--S tests give the probability that the respective field and
cluster $p$(Field/Cluster) samples are {\em not} drawn from the same continuous $h$ distributions for Type I
and Type $\rm I_o$ galaxies and $T$ distributions for Type $\rm II_o/III_o$ galaxies. Results are also shown
for when the sky is over- and undersubtracted by $\pm1\sigma$. We find no environmental dependence on
either the scalelength $h$ of Type I and Type $\rm I_o$ galaxies or the break strength $T$ of Type $\rm
II_o/III_o$ galaxies.}
\begin{tabular}{lccc}
\hline
\hline
{}				&\multicolumn{3}{c}{$p$(Field/Cluster)}				\\
\cline{2-4}
{Sky subtraction}		&{Under ($-1\sigma$)}	&{Nominal}	&{Over ($+1\sigma$)}	\\
\hline
{Type I ($h$)}			&{}			&{}		&{}			\\
{DTM}				&{$0.078$}		&{$0.101$}	&{$0.167$}		\\
{AAS}				&{$0.825$}		&{$0.761$}	&{$0.865$}		\\
{MEG}				&{$0.952$}		&{$0.919$}	&{$0.921$}
\vspace{2mm}											\\
{Type $\rm I_o$ ($h$)}		&{}			&{}		&{}			\\
{DTM}				&{$0.096$}		&{$0.146$}	&{$0.186$}		\\
{AAS}				&{$0.545$}		&{$0.545$}	&{$0.859$}		\\
{MEG}				&{$0.720$}		&{$0.716$}	&{$0.629$}
\vspace{2mm}											\\
{Type $\rm II_o/III_o$ ($T$)}	&{}			&{}		&{}			\\
{DTM}				&{$0.165$}		&{$0.038$}	&{$0.122$}		\\
{AAS}				&{$0.797$}		&{$0.908$}	&{$0.795$}		\\
{MEG}				&{$0.180$}		&{$0.307$}	&{$0.563$}		\\
\hline
\hline
\end{tabular}
\end{minipage}
\end{table}

\section[]{Conclusions}

\label{Conclusions}

We present an analysis of $V$-band radial surface brightness profiles for spiral galaxies from the field and
cluster environments using {\em HST}/ACS imaging and data from STAGES. Using a large, mass-limited ($M_* >
10^9 \,\rm M_\odot$), visually classified sample of $\sim330$ field and cluster spiral galaxies, we assess
the effect of the galaxy environment on the radial surface brightness $\mu$ profile in the outer stellar
disc ($\mu > 24 \,{\rm mag\,arcsec}^{-2}$).

We classify our spiral galaxies according to $\mu$ break features in their outer disc. If the galaxy has no
break in this $\mu$ range then the galaxy has a simple exponential profile in the outer disc and is
classified as Type $\rm I_o$. Galaxies that have a simple exponential profile across the length of their
disc component (Type I, a subset of Type $\rm I_o$ ), are also identified. However, if the galaxy exhibits
a broken exponential in the outer disc it is classified as either Type $\rm II_o$ for a down-bending break
(outer disc truncation) or Type $\rm III_o$ for an up-bending break (outer disc antitruncation).

The frequency of outer disc profile types (Type $\rm I_o$, $\rm II_o$, and $\rm III_o$) is approximately
the same in both the field and cluster environment. For both field and cluster spirals, $\sim50$ per cent
have a simple exponential profile in the outer stellar disc (Type $\rm I_o$), $\sim10$ per cent exhibit a
down-bending break (outer disc truncation, Type $\rm II_o$), and $\sim40$ per cent exhibit an up-bending
break (outer disc antitruncation, Type $\rm III_o$). These results imply that the shape of the outer disc
surface brightness profile is not dependent on the galaxy environment. These results agree for break
classifications performed by three independent assessors. We stress that due to our profile
classification being based on breaks with $\mu_{\rm brk} > 24 \,{\rm mag\,arcsec}^{-2}$, our profile type
fractions do not necessarily need to agree with those of previous works \cite[e.g.][]{Pohlen_Trujillo:2006}.

We also find that the morphological mix (Sa, Sb, Sc, Sd) of the different outer disc profile types is the
approximately same in the field and cluster environment. However, we do find a dependence of the shape
of the $\mu$ profile in the outer stellar disc on the Hubble type. Outer disc truncations are slightly
more frequent in later Hubble types while the outer disc antitruncations are slightly more frequent in
earlier Hubble types. The same dependence is observed in both the field and the cluster environments. This
result is in qualitative agreement with that of \cite{Pohlen_Trujillo:2006} who find a similar correlation
using galaxies from SDSS (classification based on entire disc component so direct comparisons cannot
be made).

For galaxies with a purely exponential outer stellar disc (Type $\rm I_o$, $\sim 50$ per cent), we find no
evidence to suggest an environmental dependence on the outer disc scalelength $h_{\rm out}$. We also find
no evidence for an environmental dependence on the scalelength $h$ for galaxies that have a pure exponential
profile across the length of their disc component (Type I, a subset of Type $\rm I_o$). For galaxies with a
broken exponential in their outer stellar disc, either down-bending (Type $\rm II_o$, outer disc truncation,
$\sim 10$ per cent) or up-bending (Type $\rm III_o$, outer disc antitruncation, $\sim 40$ per cent), we
measure the break strength as the outer-to-inner scalelength ratio
${\rm log}_{10}\,h_{\rm out}/h_{\rm in}$. We also find no evidence to suggest an environmental dependence on
this break strength. We conclude that there is no evidence to suggest the stellar distribution in the outer
stellar disc is affected by the galaxy environment for these galaxies. These results have been shown to be
robust to the sky subtraction error and to agree for break classifications performed by three independent
assessors. This work is in qualitative agreement with the work of \cite{Pohlen_Trujillo:2006} who come to a
similar conclusion for a sample of $\sim90$ spiral galaxies from SDSS (classification based on entire disc
component).

We also find that for galaxies with small effective radii ($r_e < 3\,{\rm kpc}$), Type $\rm II_o$ profiles
(outer disc truncations) are rare in both the field and cluster environments. In contrast Type $\rm III_o$
(outer disc antitruncations) are commonplace.

Our results suggest that the galaxy environment has no effect on the stellar distribution in the outer
stellar disc of spiral galaxies from the field out to the intermediate densities of the A901/2 clusters.
This implies that the origin of broken exponentials is related to an internal mechanism (e.g. star formation
threshold, resonance phenomenon) or minor mergers. Our results are consistent with previous work carried out
on the effect of the galaxy environment on disc features in the STAGES survey. \cite{Marinova_etal:2009} find
that the optical fraction of bars among disc galaxies does not show evidence for any strong variation between
the field and the A901/2 clusters, suggesting that the mass redistribution associated with bar
formation within galactic discs is not a strong function of environment from the field out to intermediate
densities. However, it is important to note that STAGES only covers an intermediate density environment
(projected galaxy number density up to $\sim1600\,{\rm gal\,Mpc^{-3}}$, \citealt{Heiderman_etal:2009}) and
not a high density environment (e.g. the COMA cluster, $\sim 10^4\,{\rm gal\,Mpc^{-3}}$,
\citealt{The_White:1986}). Therefore, it is important to investigate the effect of the environment on the
outer disc structure at higher densities (e.g. the COMA cluster).

The results presented here are for one survey field and one multicluster complex: therefore it is also
important to investigate the environmental dependence of the outer disc structure of spiral galaxies in
different surveys and across a wide redshift range. The results of such studies will enable the evolution of
the outer disc to be investigated in different galaxy environments. Also the investigation into the origin
of broken exponentials will depend on multiwavelength data being used to create colour profiles and stellar
mass distributions. This will allow an assessment of whether the origin of broken exponentials is due to the
distribution of stellar mass or a radial change in the age of the stellar population.

\section[]{Acknowledgements}

The support for STAGES was provided by NASA through GO-10395 from STScI operated by AURA under NAS5-26555.
DTM and BH were supported by STFC. MEG was supported by an STFC Advanced Fellowship. SJ acknowledges
support from the National Aeronautics and Space Administration (NASA) LTSA grant NAG5-13063, NSF grant
AST-0607748, and {\em HST} grants GO-11082  from STScI, which is operated by AURA, Inc., for NASA, under
NAS5-26555. We would also like to thank Ignacio Trujillo for useful discussions.


\bibliographystyle{mn2e} \bibliography{DTM_bibtex} \bsp

\begin{thebibliography}{}

\bibitem[\protect\citeauthoryear{{Azzollini}, {Trujillo} \&
  {Beckman}}{{Azzollini} et~al.}{2008}]{Azzollini_etal:2008}
{Azzollini} R.,  {Trujillo} I.,    {Beckman} J.~E.,  2008, \apj, 684, 1026

\bibitem[\protect\citeauthoryear{{Bakos}, {Trujillo} \& {Pohlen}}{{Bakos}
  et~al.}{2008}]{Bakos_etal:2008}
{Bakos} J.,  {Trujillo} I.,    {Pohlen} M.,  2008, \apjl, 683, L103

\bibitem[\protect\citeauthoryear{{Bland-Hawthorn}, {Vlaji{\'c}}, {Freeman} \&
  {Draine}}{{Bland-Hawthorn} et~al.}{2005}]{BlandHawthorn_etal:2005}
{Bland-Hawthorn} J.,  {Vlaji{\'c}} M.,  {Freeman} K.~C.,    {Draine} B.~T.,
  2005, \apj, 629, 239

\bibitem[\protect\citeauthoryear{{Borch}, {Meisenheimer}, {Bell}, {Rix},
  {Wolf}, {Dye}, {Kleinheinrich}, {Kovacs} \& {Wisotzki}}{{Borch}
  et~al.}{2006}]{Borch_etal:2006}
{Borch} A.,  {Meisenheimer} K.,  {Bell} E.~F.,  {Rix} H.-W.,  {Wolf} C.,  {Dye}
  S.,  {Kleinheinrich} M.,  {Kovacs} Z.,    {Wisotzki} L.,  2006, \aap, 453,
  869

\bibitem[\protect\citeauthoryear{{Bournaud}, {Elmegreen} \&
  {Elmegreen}}{{Bournaud} et~al.}{2007}]{Bournaud_etal:2007}
{Bournaud} F.,  {Elmegreen} B.~G.,    {Elmegreen} D.~M.,  2007, \apj, 670, 237

\bibitem[\protect\citeauthoryear{{Buitrago}, {Trujillo}, {Conselice},
  {Bouwens}, {Dickinson} \& {Yan}}{{Buitrago}
  et~al.}{2008}]{Buitrago_etal:2008}
{Buitrago} F.,  {Trujillo} I.,  {Conselice} C.~J.,  {Bouwens} R.~J.,
  {Dickinson} M.,    {Yan} H.,  2008, \apjl, 687, L61

\bibitem[\protect\citeauthoryear{{de Vaucouleurs}}{{de
  Vaucouleurs}}{1959}]{deVaucouleurs:1959}
{de Vaucouleurs} G.,  1959, Handbuch der Physik, 53, 311

\bibitem[\protect\citeauthoryear{{Debattista}, {Mayer}, {Carollo}, {Moore},
  {Wadsley} \& {Quinn}}{{Debattista} et~al.}{2006}]{Debattista_etal:2006}
{Debattista} V.~P.,  {Mayer} L.,  {Carollo} C.~M.,  {Moore} B.,  {Wadsley} J.,
    {Quinn} T.,  2006, \apj, 645, 209

\bibitem[\protect\citeauthoryear{{Elmegreen} \& {Parravano}}{{Elmegreen} \&
  {Parravano}}{1994}]{Elmegreen_Parravano:1994}
{Elmegreen} B.~G.,  {Parravano} A.,  1994, \apjl, 435, L121+

\bibitem[\protect\citeauthoryear{{Erwin}, {Beckman} \& {Pohlen}}{{Erwin}
  et~al.}{2005}]{Erwin_etal:2005}
{Erwin} P.,  {Beckman} J.~E.,    {Pohlen} M.,  2005, \apjl, 626, L81

\bibitem[\protect\citeauthoryear{{Ferguson}, {Irwin}, {Chapman}, {Ibata},
  {Lewis} \& {Tanvir}}{{Ferguson} et~al.}{2007}]{Ferguson_etal:2007}
{Ferguson} A.,  {Irwin} M.,  {Chapman} S.,  {Ibata} R.,  {Lewis} G.,
  {Tanvir} N.,  2007, {Resolving the Stellar Outskirts of M31 and M33}.
pp 239--+

\bibitem[\protect\citeauthoryear{{Foyle}, {Courteau} \& {Thacker}}{{Foyle}
  et~al.}{2008}]{Foyle_etal:2008}
{Foyle} K.,  {Courteau} S.,    {Thacker} R.~J.,  2008, \mnras, 386, 1821

\bibitem[\protect\citeauthoryear{{Freeman}}{{Freeman}}{1970}]{Freeman:1970}
{Freeman} K.~C.,  1970, \apj, 160, 811

\bibitem[\protect\citeauthoryear{{Giavalisco}, {Ferguson}, {Koekemoer},
  {Dickinson} \& {et al.}}{{Giavalisco} et~al.}{2004}]{Giavalisco_etal:2004}
{Giavalisco} M.,  {Ferguson} H.~C.,  {Koekemoer} A.~M.,  {Dickinson} M.,    {et
  al.} 2004, \apjl, 600, L93

\bibitem[\protect\citeauthoryear{{Gray}, {Wolf}, {Barden}, {Peng},
  {H{\"a}u{\ss}ler}, {Bell}, {McIntosh} \& {et al.}}{{Gray}
  et~al.}{2009}]{Gray_etal:2009}
{Gray} M.~E.,  {Wolf} C.,  {Barden} M.,  {Peng} C.~Y.,  {H{\"a}u{\ss}ler} B.,
  {Bell} E.~F.,  {McIntosh} D.~H.,    {et al.} 2009, \mnras, 393, 1275

\bibitem[\protect\citeauthoryear{{Gunn} \& {Gott}}{{Gunn} \&
  {Gott}}{1972}]{Gunn_Gott:1972}
{Gunn} J.~E.,  {Gott} J.~R.~I.,  1972, \apj, 176, 1

\bibitem[\protect\citeauthoryear{{Heiderman}, {Jogee}, {Marinova}, {van
  Kampen}, {Barden}, {Peng}, {Heymans}, {Gray} \& {et al.}}{{Heiderman}
  et~al.}{2009}]{Heiderman_etal:2009}
{Heiderman} A.,  {Jogee} S.,  {Marinova} I.,  {van Kampen} E.,  {Barden} M.,
  {Peng} C.~Y.,  {Heymans} C.,  {Gray} M.~E.,    {et al.} 2009, \apj, 705, 1433

\bibitem[\protect\citeauthoryear{{Ibata}, {Chapman}, {Ferguson}, {Lewis},
  {Irwin} \& {Tanvir}}{{Ibata} et~al.}{2005}]{Ibata_etal:2005}
{Ibata} R.,  {Chapman} S.,  {Ferguson} A.~M.~N.,  {Lewis} G.,  {Irwin} M.,
  {Tanvir} N.,  2005, \apj, 634, 287

\bibitem[\protect\citeauthoryear{{Jedrzejewski}}{{Jedrzejewski}}{1987}]{Jedrze%
jewski:1987}
{Jedrzejewski} R.~I.,  1987, \mnras, 226, 747

\bibitem[\protect\citeauthoryear{{Kennicutt}
  Jr.}{{Kennicutt}}{1989}]{Kennicutt:1989}
{Kennicutt} Jr. R.~C.,  1989, \apj, 344, 685

\bibitem[\protect\citeauthoryear{{Maltby}, {Arag{\'o}n-Salamanca}, {Gray},
  {Barden}, {H{\"a}u{\ss}ler}, {Wolf}, {Peng}, {Jahnke}, {McIntosh}, {B{\"o}hm}
  \& {van Kampen}}{{Maltby} et~al.}{2010}]{Maltby_etal:2010}
{Maltby} D.~T.,  {Arag{\'o}n-Salamanca} A.,  {Gray} M.~E.,  {Barden} M.,
  {H{\"a}u{\ss}ler} B.,  {Wolf} C.,  {Peng} C.~Y.,  {Jahnke} K.,  {McIntosh}
  D.~H.,  {B{\"o}hm} A.,    {van Kampen} E.,  2010, \mnras, 402, 282

\bibitem[\protect\citeauthoryear{{Marinova}, {Jogee}, {Heiderman}, {Barazza},
  {Gray}, {Barden}, {Wolf}, {Peng} \& {et al.}}{{Marinova}
  et~al.}{2009}]{Marinova_etal:2009}
{Marinova} I.,  {Jogee} S.,  {Heiderman} A.,  {Barazza} F.~D.,  {Gray} M.~E.,
  {Barden} M.,  {Wolf} C.,  {Peng} C.~Y.,    {et al.} 2009, \apj, 698, 1639

\bibitem[\protect\citeauthoryear{{Moore}, {Katz}, {Lake}, {Dressler} \&
  {Oemler}}{{Moore} et~al.}{1996}]{Moore_etal:1996}
{Moore} B.,  {Katz} N.,  {Lake} G.,  {Dressler} A.,    {Oemler} A.,  1996,
  \nat, 379, 613

\bibitem[\protect\citeauthoryear{{Okamoto} \& {Nagashima}}{{Okamoto} \&
  {Nagashima}}{2004}]{Okamoto_Nagashima:2004}
{Okamoto} T.,  {Nagashima} M.,  2004, in {Diaferio} A.,  ed., IAU Colloq. 195:
  Outskirts of Galaxy Clusters: Intense Life in the Suburbs {The roles of
  ram-pressure stripping and minor mergers in the evolution of galaxies}.
pp 534--538

\bibitem[\protect\citeauthoryear{{Patterson}}{{Patterson}}{1940}]{Patterson:19%
40}
{Patterson} F.~S.,  1940, Harvard College Observatory Bulletin, 914, 9

\bibitem[\protect\citeauthoryear{{Peng}, {Ho}, {Impey} \& {Rix}}{{Peng}
  et~al.}{2002}]{Peng_etal:2002}
{Peng} C.~Y.,  {Ho} L.~C.,  {Impey} C.~D.,    {Rix} H.-W.,  2002, \aj, 124, 266

\bibitem[\protect\citeauthoryear{{P{\'e}rez}}{{P{\'e}rez}}{2004}]{Perez:2004}
{P{\'e}rez} I.,  2004, \aap, 427, L17

\bibitem[\protect\citeauthoryear{{Pohlen}, {Dettmar}, {L{\"u}tticke} \&
  {Aronica}}{{Pohlen} et~al.}{2002}]{Pohlen_etal:2002}
{Pohlen} M.,  {Dettmar} R.,  {L{\"u}tticke} R.,    {Aronica} G.,  2002, \aap,
  392, 807

\bibitem[\protect\citeauthoryear{{Pohlen} \& {Trujillo}}{{Pohlen} \&
  {Trujillo}}{2006}]{Pohlen_Trujillo:2006}
{Pohlen} M.,  {Trujillo} I.,  2006, \aap, 454, 759

\bibitem[\protect\citeauthoryear{{Pohlen}, {Zaroubi}, {Peletier} \&
  {Dettmar}}{{Pohlen} et~al.}{2007}]{Pohlen_etal:2007}
{Pohlen} M.,  {Zaroubi} S.,  {Peletier} R.~F.,    {Dettmar} R.,  2007, \mnras,
  378, 594

\bibitem[\protect\citeauthoryear{{Ro{\v s}kar}, {Debattista}, {Quinn},
  {Stinson} \& {Wadsley}}{{Ro{\v s}kar} et~al.}{2008}]{Roskar_etal:2008b}
{Ro{\v s}kar} R.,  {Debattista} V.~P.,  {Quinn} T.~R.,  {Stinson} G.~S.,
  {Wadsley} J.,  2008, \apjl, 684, L79

\bibitem[\protect\citeauthoryear{{Ro{\v s}kar}, {Debattista}, {Stinson},
  {Quinn}, {Kaufmann} \& {Wadsley}}{{Ro{\v s}kar}
  et~al.}{2008}]{Roskar_etal:2008a}
{Ro{\v s}kar} R.,  {Debattista} V.~P.,  {Stinson} G.~S.,  {Quinn} T.~R.,
  {Kaufmann} T.,    {Wadsley} J.,  2008, \apjl, 675, L65

\bibitem[\protect\citeauthoryear{{Schaye}}{{Schaye}}{2004}]{Schaye:2004}
{Schaye} J.,  2004, \apj, 609, 667

\bibitem[\protect\citeauthoryear{{S{\'e}rsic}}{{S{\'e}rsic}}{1968}]{Sersic:196%
8}
{S{\'e}rsic} J.~L.,  1968, {Atlas de galaxias australes}.
Cordoba, Argentina: Observatorio Astronomico, 1968

\bibitem[\protect\citeauthoryear{{The} \& {White}}{{The} \&
  {White}}{1986}]{The_White:1986}
{The} L.~S.,  {White} S.~D.~M.,  1986, \aj, 92, 1248

\bibitem[\protect\citeauthoryear{{Trujillo} \& {Pohlen}}{{Trujillo} \&
  {Pohlen}}{2005}]{Trujillo_Pohlen:2005}
{Trujillo} I.,  {Pohlen} M.,  2005, \apjl, 630, L17

\bibitem[\protect\citeauthoryear{{van der Kruit}}{{van der
  Kruit}}{1979}]{vanderKruit:1979}
{van der Kruit} P.~C.,  1979, \aaps, 38, 15

\bibitem[\protect\citeauthoryear{{van der Kruit}}{{van der
  Kruit}}{2007}]{vanderKruit:2007}
{van der Kruit} P.~C.,  2007, \aap, 466, 883

\bibitem[\protect\citeauthoryear{{Wolf}, {Arag{\'o}n-Salamanca}, {Balogh},
  {Barden}, {Bell}, {Gray}, {Peng} \& {et al.}}{{Wolf}
  et~al.}{2009}]{Wolf_etal:2009}
{Wolf} C.,  {Arag{\'o}n-Salamanca} A.,  {Balogh} M.,  {Barden} M.,  {Bell}
  E.~F.,  {Gray} M.~E.,  {Peng} C.~Y.,    {et al.} 2009, \mnras, 393, 1302

\bibitem[\protect\citeauthoryear{{Wolf}, {Hildebrandt}, {Taylor} \&
  {Meisenheimer}}{{Wolf} et~al.}{2008}]{Wolf_etal:2008}
{Wolf} C.,  {Hildebrandt} H.,  {Taylor} E.~N.,    {Meisenheimer} K.,  2008,
  \aap, 492, 933

\bibitem[\protect\citeauthoryear{{Wolf}, {Meisenheimer}, {Kleinheinrich},
  {Borch}, {Dye}, {Gray}, {Wisotzki}, {Bell}, {Rix}, {Cimatti}, {Hasinger} \&
  {Szokoly}}{{Wolf} et~al.}{2004}]{Wolf_etal:2004}
{Wolf} C.,  {Meisenheimer} K.,  {Kleinheinrich} M.,  {Borch} A.,  {Dye} S.,
  {Gray} M.,  {Wisotzki} L.,  {Bell} E.~F.,  {Rix} H.-W.,  {Cimatti} A.,
  {Hasinger} G.,    {Szokoly} G.,  2004, \aap, 421, 913

\bibitem[\protect\citeauthoryear{{Wolf}, {Meisenheimer}, {Rix}, {Borch}, {Dye}
  \& {Kleinheinrich}}{{Wolf} et~al.}{2003}]{Wolf_etal:2003}
{Wolf} C.,  {Meisenheimer} K.,  {Rix} H.-W.,  {Borch} A.,  {Dye} S.,
  {Kleinheinrich} M.,  2003, \aap, 401, 73

\bibitem[\protect\citeauthoryear{{York}, {Adelman}, {Anderson} Jr., {Anderson},
  {Annis}, {Bahcall}, {Bakken} \& {et al.}}{{York}
  et~al.}{2000}]{York_etal:2000}
{York} D.~G.,  {Adelman} J.,  {Anderson} Jr. J.~E.,  {Anderson} S.~F.,  {Annis}
  J.,  {Bahcall} N.~A.,  {Bakken} J.~A.,    {et al.} 2000, \aj, 120, 1579

\bibitem[\protect\citeauthoryear{{Younger}, {Cox}, {Seth} \&
  {Hernquist}}{{Younger} et~al.}{2007}]{Younger_etal:2007}
{Younger} J.~D.,  {Cox} T.~J.,  {Seth} A.~C.,    {Hernquist} L.,  2007, \apj,
  670, 269

\end{thebibliography}

\label{lastpage}

\end{document}